\documentclass{article}
\usepackage{authblk}
\usepackage{graphicx}
\usepackage{amsmath}
\usepackage{amsfonts}
\usepackage{xcolor}
\usepackage{float}
\usepackage{color,soul}
\usepackage[bookmarks=false,hidelinks]{hyperref}
\usepackage{comment}
\usepackage{csvsimple}
\usepackage{booktabs}
\usepackage{caption}
\usepackage{amsmath}
\usepackage{bbm}
\usepackage{natbib}
\usepackage[customcolors]{hf-tikz}
\usepackage[linesnumbered,ruled,vlined]{algorithm2e} 
\usepackage{todonotes}
\usepackage{authblk}
\usepackage{placeins} 
\usepackage{tabularx}

\usepackage{soul}

\hfsetfillcolor{green!10}
\newcounter{suppfigure}

\title{Evaluating Gender Wage Inequality in Academia using Causal Inference Methods for Observational Data}
\author[1]{Zihan Zhang}
\author[1]{Jan Hannig}
\affil[1]{Department of Statistics and Operations Research, University of North Carolina at Chapel Hill, Chapel Hill, NC, USA}

\date{\today}

\begin{document}

\maketitle

\newpage

\newpage

\begin{abstract}

Observational studies often present challenges for causal inference due to confounding and heterogeneity. In this paper, we illustrate how modern causal inference methods can be applied to large-scale academic salary data. Using records from 12,039 tenure-track faculty in the University of North Carolina system, linked with bibliometric indicators and institutional classifications, we estimate the causal effect of gender on faculty salaries. Our analysis combines propensity score matching with causal forests to adjust for rank, discipline, research productivity, and career experience. Results indicate that female faculty earn approximately 6\% less than comparable male colleagues, with variation in the gap across career stages and levels of research productivity. This case study demonstrates how causal inference methods for observational data can provide insight into structural disparities in complex social systems.

\end{abstract}

\section{Introduction}\label{sec1}

Understanding and explaining the structure and persistence of the gender wage gap remains a central question in labor economics and the social sciences. Since the foundational work of \citet{blinder1973wage} and \citet{oaxaca1973male}, the literature has sought to distinguish disparities attributable to productive characteristics from those arising through labor-market institutions. The academic labor market, which prides itself on meritocracy and performance-based advancement, presents a particularly puzzling case: despite controlling for rank, field, and research output, standard wage equations consistently reveal lower earnings for women faculty relative to their male counterparts \citep{ginther2004women, ceci2014women, monroe2008gender}, yet this gap has rarely been examined using causal estimators. These disparities persist despite the growing representation of women in academia and widespread institutional efforts to promote equity in hiring and advancement \citep{blau2017gender,BLS2023}.

Two main perspectives have shaped scholarly understanding of gender pay gaps in academia. One attributes disparities to structural factors: women are more likely to occupy lower-paid positions—such as assistant professorships, non-tenure-track roles, or teaching-intensive appointments—which naturally depresses their average earnings \citep{ginther2004women, monroe2008gender}. Another perspective emphasizes differences in human capital, suggesting that salary disparities reflect variation in research productivity, often measured through publication counts or citation impact \citep{ceci2014women, kim2024gendergap}. While some studies—such as \citet{ceci2014women}—report relatively small pay differences after accounting for productivity, their work continues to treat scholarly output as the primary metric of academic value. In practice, however, it is difficult to disentangle these explanations, as rank, research output, and compensation typically evolve in tandem over the course of a faculty member’s career.

Much of the existing empirical literature relies on associational methods, such as Oaxaca–Blinder decompositions or pooled regressions based on cross-sectional data. These approaches are limited in that they cannot estimate counterfactual salary outcomes, nor can they fully account for unobserved confounding—such as career interruptions or uneven service burdens—that may influence both productivity and compensation. While prior work has documented the existence of gender pay disparities in academia, there remains limited causal evidence on how these gaps vary by institutional type, career stage, or research performance. These gaps are central to identifying equity interventions.

In this study, we address these limitations by applying modern causal inference tools to examine the gender wage gap among professors in the North Carolina public university system. We merge faculty salary records from all sixteen campuses in the University of North Carolina system—which are legally required to disclose annual salaries—with academic performance data from Google Scholar and gender labels inferred via genderize.io. Our dataset covers salary information from 2022 and includes 12,039 faculty, with detailed information on academic rank, department, institutional affiliation, years of experience, and research productivity, measured using the log-transformed i10-index. This rich dataset enables us to move beyond descriptive associations and address core challenges in observational causal inference, such as confounding, covariate imbalance, and treatment effect heterogeneity, by estimating the causal effect of gender on log-transformed salary outcomes. We also explore treatment effect heterogeneity by examining how the gender wage gap varies across career stages and levels of research performance.

Our study contributes to the literature on gender inequality and academic labor markets in three key ways. First, we provide a causal estimate of the gender wage gap across a comprehensive public university system, offering a more generalizable view of pay disparities than analyses limited to a single institution or field. Second, we move beyond average treatment effects by applying causal forests to uncover patterns of heterogeneity in the gender wage gap. This method reveals how the magnitude of pay disparities varies across career stages and levels of research productivity, offering new insights into where interventions may be most needed. Third, we release a reproducible workflow that integrates public salary disclosures, bibliometric data from Google Scholar, and modern causal inference techniques. This framework can be readily adapted for equity audits in other academic systems.

The remainder of the paper proceeds as follows. Section~\ref{sec2} describes the dataset and variable construction. Section~\ref{sec3} outlines the methodological framework, including both parametric and non-parametric approaches. Section~\ref{sec4} presents the empirical findings. Section~\ref{sec5} concludes with implications and future directions.

\section{Data and Variables}\label{sec2}
To investigate the relationship between gender and faculty salary, we compiled a large database of employee salaries from 2022 of all sixteen institutions in North Carolina public university system. These sixteen universities within the one university system share the same governance structure and operate under the same policy orientation including faculty hiring and promotion \citep{UNCpolicy2023}. The data used in this study was obtained from the University of North Carolina Digital Measures website \citep{UNCsalary2022}. The dataset contains 47,405 entries, with information on full names, age, initial hired date, rank, department, and the salaries of all employees, from faculty members to administrative and support staff, affiliated with one of the institutions. We removed non-faculty employees (such as department staff, facilities employees, athletic coaches, etc.) from the data based on the employee home department and a column that indicates job functions. Also, the professors with less than $27,000$ yearly salary were eliminated \citep{lu2024gender} as these are likely clerical errors. In total, seven observations were dropped from the dataset. The resulting dataset contains 12,039 entries, focusing on tenure track faculty (assistant professor, associate professor, and full professor) in all institutions from North Carolina public university system.

\begin{figure}[H] 
    \centering
    \begin{minipage}[t]{0.48\textwidth}
        \centering
        \includegraphics[width=\textwidth]{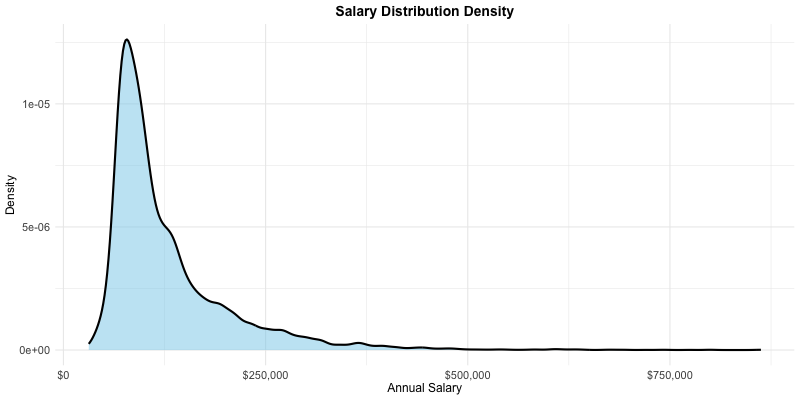}
        \caption{Annual salary distribution of faculties in NC public university system in year 2022.}
        \label{fig:salary_distribution}
    \end{minipage}
    \hfill
    \begin{minipage}[t]{0.48\textwidth}
        \centering
        \includegraphics[width=\textwidth]{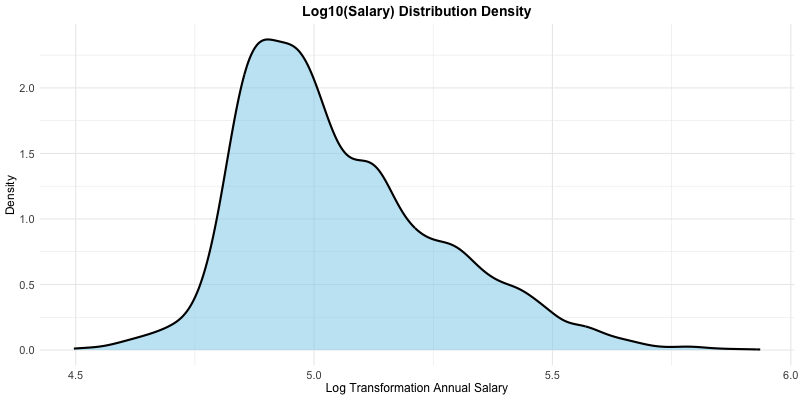}
        \caption{Log-transformed salary distribution of faculties in NC public university system in year 2022.}
        \label{fig:log_salary_distribution}
    \end{minipage}
\end{figure}

Our study focuses on examining differences in faculty salary across gender. The primary outcome variable is annual faculty salary. This variable was directly generated from the raw dataset. In our cleaned dataset, the average yearly salary is \$128,000. We use a log\textsubscript{10} transformation for our analysis because salary distributions tend to be right-skewed, see Figure \ref{fig:salary_distribution}, and the logarithmic transformation normalizes the data and reduces the influence of extreme values, Figure \ref{fig:log_salary_distribution}. Additionally, this transformation allows us to interpret the results in terms of percentage change rather than absolute dollar amounts, which is more meaningful when analyzing salary differentials across different faculty ranks and disciplines. Importantly, faculty appointments in the North Carolina public university system are explicitly categorized into either nine- or twelve-month “service periods” \citep{unc_facultyaffairs_service_periods}. Most teaching and research faculties hold standard nine-month academic-year contracts, making their base salaries directly comparable. For the relatively few twelve-month appointments (e.g., in medical schools), any salary differences reflect discipline-specific norms and are absorbed by our department fixed effects. This variable contains 12,039 observations with no missing values.

Since our raw dataset does not contain faculty gender information, we followed established methodologies from the literature to infer gender based on first names \citep{kim2022gendered}. We employed the Genderize API \footnote{\url{https://genderize.io/}} to assign gender labels to our dataset. The API returns four key variables for each query: the predicted gender classification, the confidence score of the prediction ranging from 0--100\%, the number of samples used for the prediction, and the queried first name. To ensure high accuracy in our gender assignments, we implemented a two-stage process. First, we used the Genderize API to assign gender automatically for cases with confidence scores greater than or equal to 60\%. Second, we performed manual verification for the remaining cases, including names with confidence scores below 60\%, cases where gender remains undetermined, and uncommon or culturally specific names. The number of cases dealt with each methods are shown in Table~\ref{tab:gender_distribution}. This hybrid approach, combining algorithmic prediction with manual verification, follows best practices established in recent literature \citep{vanhelene2024inferring} and helps reduce potential biases in automated gender inference systems. The manual verification process involved considering faculty websites, professional profiles, and institutional directories to determine gender with certainty. After this hybrid solution, there is no missing data for gender. Gender is recorded as a binary variable indicating whether a faculty member is classified as female or male. In this dataset, 45.5\% of the entries correspond to females, while 54.5\% represent males. Although gender is not a manipulable variable, it is not randomly distributed across faculty in our dataset. Instead, its distribution varies systematically across institutional, disciplinary, and career-related characteristics, reflecting differences in how faculty are distributed across positions.

In addition to salary and gender, we included a set of observed variables that capture institutional, disciplinary, and career-related characteristics of faculty, including working years, academic title, university classification, department, and academic performance.

The first variable we considered is length of employment. To consider career progression effects on salary, we calculated working years as the difference between 2022 and each faculty member's initial hire date. The average working years in our sample is $14.6$ years (SD = $9.82$). The variable \texttt{working years} has no missing values.

Next, we also included \texttt{title} in our analysis, since title information helps control for career stage, which strongly influences salary levels \citep{griswold1996political}. This information is provided in the raw dataset and has no missing values. We removed non-faculty and non-tenure track employees and categorized faculty titles into three groups: Assistant Professor (35.1\%), Associate Professor (31.0\%), and Professor (34.0\%).

To account for effects related to the various missions of the 16 universities within the UNC system, we used variable \texttt{university classification}. Following Carnegie Classification of Institutions of Higher Education\footnote{\url{https://carnegieclassifications.acenet.edu/}}, which is a widely used framework for categorizing colleges and universities in the United States, we categorized these institutions into four types: bachelor (3 institutions), master (5 institutions), DRU(H) (6 institutions), and DU/VA (2 institutions). The first category is Bachelor's Colleges and Universities (bachelor), in which the institutions award bachelor's degrees to at least 50\% of their graduates, while awarding fewer than 50 master's degrees and fewer than 20 doctoral degrees annually. The second is Master's Colleges and Universities (master). These institutions award at least 50 master's degrees annually, while awarding fewer than 20 doctoral degrees annually. For Doctoral/Research Universities (DRU(H)), the universities award at least 20 research/scholarship doctoral degrees annually (excluding professional degrees such as MD, JD, PharmD). The last category is Doctoral Universities or Very High Research Activity (DU/VA), where the institutions are characterized by very high levels of research activity and a significant output of doctoral degrees in diverse fields, often referred to as R1 universities. For our analysis, we combined bachelor's and master's institutions into a single category, as both types of institutions share similar characteristics in terms of their primary focus on teaching rather than research \citep{henderson2007scholarship}. Across the UNC system: 47.4\% are affiliated with R1 universities (DU/VA), 38.6\% with R2 universities (DRU(H)), and 14.0\% with primarily undergraduate or master’s institutions (bachelor/master).

Controlling for faculty's home department is crucial because academic disciplines significantly influence salary levels, with substantial variations observed across fields \citep{toutkoushian2007interaction}. We categorized departments into six broad fields: Business (6.5\%), Technology and Engineering (15.3\%), Arts and Humanities (18.1\%), Medical and Health Sciences (27.2\%), Natural Sciences (16.3\%), and Social Sciences (16.6\%). This categorization follows the UNESCO's International Standard Classification of Education (ISCED-F) framework \citep{unesco2014isced}, which provides a standardized system for classifying academic disciplines in higher education. We used Large Language Models (LLMs), specifically \texttt{ChatGPT}, for department name standardization. LLMs, based on transformer architectures \citep{vaswani2017attention}, have demonstrated superior performance in understanding semantic relationships and contextual variations in text \citep{brown2020language}. This approach enabled efficient and accurate mapping of diverse department names into our predefined categories while maintaining consistency across institutions. To ensure reliability, a random sample of ChatGPT-generated classifications was manually reviewed and confirmed.

\begin{figure}[H]
    \centering
    \begin{minipage}[t]{0.48\textwidth}
        \centering
        \includegraphics[width=\textwidth]{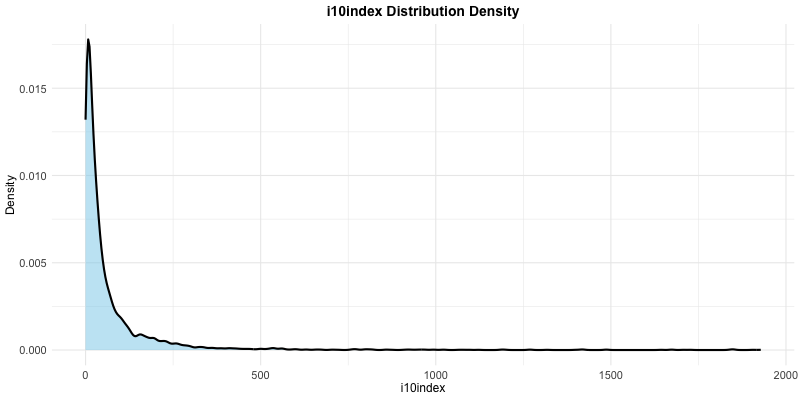}
        \caption{i10index Distribution Density}
        \label{fig:i10index_dist}
    \end{minipage}
    \hfill
    \begin{minipage}[t]{0.48\textwidth}
        \centering
        \includegraphics[width=\textwidth]{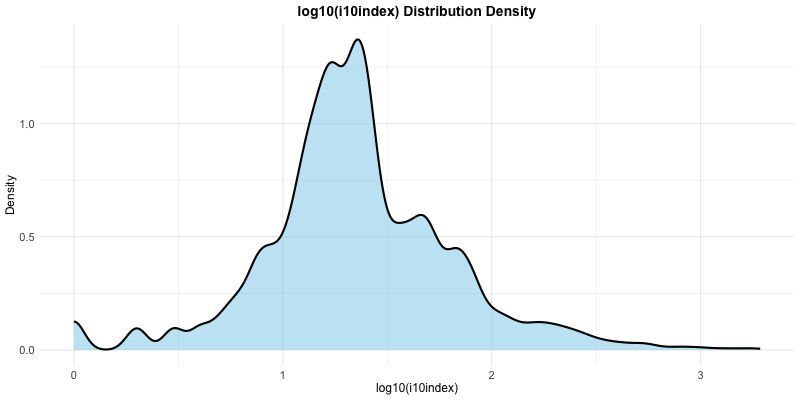}
        \caption{log10(i10index) Distribution Density}
        \label{fig:log10_i10index_dist}
    \end{minipage}
\end{figure}

Academic performance may also significantly influence faculty compensation \citep{fairweather2005beyond}. To account for this we matched faculty members to their Google Scholar profiles when available (match rate = 41.4\%) by using \texttt{Google Scholar API}. This identifier enables us to collect publication metrics and establish research productivity measures. From Google Scholar, we extracted several metrics for each faculty member: total citations (citedby), 5-year citations (citedby5y), h-index (hindex), 5-year h-index (hindex5y), i10-index (i10index), and 5-year i10-index (i10index5y). To deal with the max access limitations of \texttt{Google Scholar API}, we divided the data into smaller files combining them into one complete file at the end. In order to address the possibility of retrieving information for another faculty member with identical names rather than the intended individual, we incorporated an additional validation step using the email domain provided by the \texttt{Google Scholar API}. Specifically, we matched the email domain with the university's domain to ensure accuracy. Entries with matching domains were retained, while those without a match were excluded. However, this approach introduces a new limitation. Faculty members who have used an email address associated with a previous institution may be incorrectly excluded, as the domain would not align with their current university. To avoid the limitation caused by the email matching exclusion, we manually tested the excluded entries. Consequently, the three-step testing method provides a more reliable linkage between faculty and their profiles, ensuring higher overall accuracy. Among academic metrics from Google Scholar API, we selected \texttt{i10-index}, which measures research impact by counting publications with at least 10 citations \citep{google2011metrics}, as our primary measure of research productivity because it captures both the quantity and impact of scholarly work. We also tried other academic indicators such as h-index, but in our models the i10-index was the best metric for academic performance. We applied log\textsubscript{10} transformation to i10-index because citation metrics typically follow a highly right-skewed distribution  \citep{kim2020logtransform}, shown in Figure \ref{fig:i10index_dist}.  The log transformation helps normalizing the distribution and reduces the influence of extreme values in our analysis, see Figure \ref{fig:log10_i10index_dist}; the mean log10(i10-index) is 1.36 (SD = 0.456).  For faculties without Google Scholar ID, we imputed their log\textsubscript{10}(i10index) using the mean value of similar background faculty members who had available academic information. Using this metric helps control for research productivity's influence on salary, as higher research output and citation impact are associated with higher faculty compensation \citep{hirsch2005hindex}.

Table~\ref{tab:summary_nc_professors} in the Appendix presents the number of faculty members, their average salary, the median of their salary, and the Carnegie classification for each university, to help contextualize salary differences across institutions.

\section{Methodology}\label{sec3}
\subsection{Preliminary Analysis and Motivation for Causal Inference}
We began our analysis by examining the raw salary differences between male and female faculty members, without adjusting for any covariates. We plotted the data and calculated the gender pay gap as the percentage difference between the average salaries of male and female faculty. These descriptive comparisons provided a baseline understanding of our dataset.
However, these descriptive analyses do not imply causal relationships, as they do not account for potential confounding factors such as academic rank, years of experience, institutional affiliation, and research productivity.

To further investigate the relationship between gender and salary while accounting for observable differences, we used a standard linear regression model. Although linear regression does not rely on design-based causal inference, it serves as a useful benchmark for comparison with the causal models that follow. In particular we estimated the following model:
\begin{equation*}
\log(\text{Salary}_i) = \beta_0 + \beta_1 \cdot \text{Gender}_i + \mathbf{X}_i \boldsymbol{\beta} + \varepsilon_i,
\end{equation*}
where \( \text{Gender}_i \) is a binary indicator (1 = Female), and \( \mathbf{X}_i \) includes institutional affiliation, department, academic rank, working years, and research productivity (log$_{10}$(i10-index)). The estimated coefficient on gender in this baseline regression captures the adjusted association between gender and log salary, after controlling for observable covariates. However, as this is an observational study, the regression model assumes that all relevant confounders are correctly specified and linearly adjusted for. This approach does not account for selection bias, unobserved confounding, or treatment effect heterogeneity.

To draw valid conclusions about the causal effect of gender on salary, it is essential to use methods that account for selection bias and covariate imbalance. Therefore, we employed causal inference techniques, including both parametric and non-parametric approaches, which aim to estimate treatment effects under weaker functional form assumptions and greater model flexibility. These methods help isolate the impact of gender from other salary-determining factors while also allowing for the exploration of heterogeneous treatment effects across subgroups.

\subsection{Causal Inference Framework}
Causal inference methods aim to estimate the effect of a treatment (in this case, gender) on an outcome (salary), while addressing confounding. In our framework, gender is treated as a binary exposure variable ($Z_i$), where $Z_i = 1$ for female and $Z_i = 0$ for male, and $Y_i$ represents the observed salary. The covariate vector $X_i$ consists of title, discipline, university, working years, and performance (represented by log-transformed i10-index), capturing institutional characteristics, professional experience, and research productivity.

To clarify the relationships among gender, salary, and observed covariates, we present a conceptual directed acyclic graph (DAG) in Figure~\ref{fig:dag}. In this DAG, the observed covariates $X_i$ influence both salary and the observed distribution of gender across faculty positions. While gender itself is immutable at the individual level, the gender composition of faculty across positions is shaped by institutional processes such as hiring, promotion, and retention. As a result, variables such as academic title, discipline, institutional classification, and career stage affect the probability of being observed as male or female in the dataset, while also being directly related to salary outcomes. Under this interpretation, these variables can be viewed as confounders, as they capture structural differences that jointly determine gender composition and salary. Accordingly, we condition on a set of observed covariates $X_i$ to compare male and female faculty within similar institutional, disciplinary, and career contexts.

\begin{figure}[H]
\centering
\begin{tikzpicture}[>=stealth, node distance=2.5cm]

\node (U) at (-3,2) {University};
\node (W) at (0,2) {Working Years};
\node (D) at (3,2) {Discipline};
\node (A) at (-1.5,0.5) {Performance};
\node (T) at (1.5,0.5) {Title};
\node (G) at (-2,-1.5) {Gender};
\node (S) at (2,-1.5) {Salary};

\draw[->] (U) -- (G);
\draw[->] (U) -- (S);

\draw[->] (D) -- (G);
\draw[->] (D) -- (S);

\draw[->] (W) -- (A);
\draw[->] (W) -- (T);
\draw[->] (W) -- (G);
\draw[->] (W) -- (S);

\draw[->] (A) -- (T);
\draw[->] (A) -- (G);
\draw[->] (A) -- (S);

\draw[->] (T) -- (G);
\draw[->] (T) -- (S);

\draw[->] (G) -- (S);

\end{tikzpicture}
\caption{Conceptual DAG illustrating the relationship between gender, salary, and observed covariates. While gender is immutable at the individual level, the gender composition of faculty is shaped by institutional processes.}
\label{fig:dag}
\end{figure}

The core of causal inference lies in comparing outcomes under different treatment assignments. Specifically, we aim to estimate quantities such as the \textit{Average Treatment Effect (ATE)}, \textit{Average Treatment Effect on the Treated (ATT)}, and the \textit{Conditional Average Treatment Effect (CATE)}. ATE captures the expected difference in salary between being female and being male across the population:
\[
\text{ATE} = \mathbb{E}[Y_i(1) - Y_i(0)],
\]
where \( Y_i(1) \) and \( Y_i(0) \) represent the potential salaries for individual \( i \) if they were female or male, respectively. The Average Treatment Effect on the Treated (ATT) is defined as
\[
\text{ATT} = \mathbb{E}[Y_i(1) - Y_i(0) \mid Z_i = 1],
\]
which represents the expected effect of gender among individuals who are observed as female. The Conditional Average Treatment Effect (CATE) is defined as
\[
\text{CATE}(X_i) = \mathbb{E}[Y_i(1) - Y_i(0) \mid X_i],
\]
which captures how treatment effects vary across individuals with different observed characteristics.

These formulations inherently rely on the concept of counterfactuals: for each individual, we observe only one of the two potential outcomes---either as male or female---but not both. The unobserved outcome represents the counterfactual, and causal inference techniques are designed to estimate this missing information using observed data and plausible assumptions \citep{imbens2015causal}.

To estimate these causal effects, we employ both parametric methods (e.g., propensity score matching, IPTW, and regression adjustment with propensity score) and non-parametric methods (e.g., causal forests). Parametric approaches rely on specified functional forms and modeling assumptions, while non-parametric approaches are more flexible and allow for heterogeneous treatment effects across individuals.

To ensure valid causal estimation, we rely on three key assumptions:

\begin{enumerate}
    \item \textbf{Unconfoundedness}: Treatment assignment is independent of potential outcomes, conditional on observed covariates:
    \[
    \{Y_i(0), Y_i(1)\} \perp Z_i \mid X_i
    \]
    This assumption is satisfied automatically in randomized experiments but must be carefully justified in observational studies. We assessed balance using Standardized Mean Differences (SMDs) and love plots after covariate adjustment. To assess the robustness of our findings to potential unmeasured confounding, we conducted a sensitivity analysis using the framework of \citet{cinelli2020making}. This approach quantifies the strength of an unobserved confounder required to explain away the estimated treatment effect in terms of partial $R^2$. In particular, we report robustness values that indicate how strongly an omitted variable would need to be associated with both treatment assignment and the outcome to invalidate our conclusions.

    \item \textbf{Positivity (Overlap)}: Every individual has a non-zero probability of receiving either treatment:
    \[
    0 < P(Z_i = 1 \mid X_i) < 1
    \]
    We evaluated this by inspecting the distribution of propensity scores across treatment groups.

    \item \textbf{Stable Unit Treatment Value Assumption (SUTVA)}: The observed outcome equals the potential outcome corresponding to the treatment received:
    \[
    Y_i^{\text{obs}} = Z_i \cdot Y_i(1) + (1 - Z_i) \cdot Y_i(0)
    \]
    This also assumes that treatment is well-defined and that there is no interference between units.
\end{enumerate}

These assumptions enable identification of average treatment effects (ATE), average treatment effect on the treated (ATT), and conditional average treatment effects (CATE).

\subsection{Parametric Methods}

In this section, we implemented a set of parametric causal inference approaches to estimate the gender wage gap. Specifically, we combine propensity score–based design methods with regression adjustment to improve covariate balance and reduce residual bias. 

All approaches rely on a common propensity score specification, followed by different estimation strategies, including matching, weighting, and regression-based adjustment. This unified framework allows for consistent comparison across methods.

\paragraph{Propensity Score Specification}
We estimate the probability of being female using a logistic regression with higher-order interaction terms:

\begin{align*}
\text{logit}\big(P(\text{Gender}_i = 1 \mid X_i)\big)
=\;& \beta_0 
+ \sum_{j} \beta_{1j}\,[\text{University Code}\times \text{Department Code}]_{ij}\\
&\times \log_{10}(i10index_i) \\
&+ \sum_{k} \beta_{2k}\,(\text{Titles}_{ik}\cdot \text{Working Years}_i).
\end{align*}
where $\text{logit}(p) = \ln\left(\frac{p}{1-p}\right)$ and $[\text{University Code}\times \text{Department Code}]_{ij}$ denotes the indicator for the $j$th university-by-department combination.
Here, each $\beta_{1j}$ and $\beta_{2k}$ corresponds coefficients of dummy-coded interaction terms, reflecting the fact that both the university affiliation and academic title variables are represented using indicator variables in the model. The first captures the joint effects of institutional affiliation and research productivity, while the second reflects the interaction between academic rank and experience. Although Google Scholar ID is an indicator of academic presence and visibility, including it in the propensity score model substantially worsened post-matching covariate balance. Therefore, in the full dataset we excluded Google Scholar ID from the propensity score process to improve balance, and instead incorporated it in the outcome regression as an adjustment covariate. In the reduced subset, Google Scholar ID is constant and thus plays no role in the analysis. So, we did not include Google Scholar ID in the propensity score process. Interaction terms were selected through an iterative model development process guided by balance diagnostics, including standardized mean differences and covariate balance plots. Simpler models without these interactions resulted in poorer balance and weaker covariate overlap, underscoring the importance of capturing these complex structures in academic settings.

\paragraph{1. Propensity Score Matching (PSM) with Regression Adjustment}
After estimating the propensity scores, we performed nearest-neighbor matching with replacement, using a caliper of 0.2 standard deviations of the logit of the propensity score. We chose nearest-neighbor matching because it is a widely used and interpretable approach that pairs treated and control units with the closest estimated propensity scores, thereby producing samples that are easy to compare directly. The caliper size of 0.2 was selected following common practice in the causal inference literature, where simulation studies have shown that this threshold effectively reduces bias without discarding excessive observations \citep{austin2011optimal}.

While post-matching diagnostics indicate good marginal balance, these diagnostics do not fully capture balance in higher-order interactions \citep{austin2009balance}. Therefore, following the double-adjustment approach proposed by \citet{nguyen2017double}, we further apply regression adjustment with ordinary least squares (OLS) regression on the matched sample, incorporating the same interaction structure as specified in the propensity score model. Specifically, the model is formulated as:

\begin{align*}
\log(\text{Salary}_i) =\ & \alpha + \tau \cdot \text{Gender}_i \\
&+ \sum_{j} \gamma_{1j} \, [\text{University Code} \times \text{Department Code}]_{ij} \times  \log_{10}(i10index_i) \\
&+ \sum_{k} \gamma_{2k} \, (\text{Titles}_{ik} \times \text{Working Years}_i) + \epsilon_i,
\end{align*}
where $\tau$ represents the estimated average treatment effect of gender on the (log-transformed) salary, conditional on covariates including three-way interactions between university type, department field, and research productivity, and two-way interactions between academic rank and working years. This two-step approach, matching followed by regression adjustment, helps reduce bias due to residual imbalance and enhances robustness in the estimated gender effect.

\paragraph{2. Inverse Probability Treatment Weighting (IPTW) with Regression Adjustment}

Alternatively, following \citet{gabriel2024iptw}, we implemented inverse probability treatment weighting (IPTW) using the same propensity score model. Each observation was weighted by the inverse of its estimated treatment probability to construct a pseudo-population in which treatment assignment was independent of observed covariates. To maintain consistency with the PSM approach and further mitigate model misspecification, we applied the same regression adjustment as above on the weighted sample, including all interaction terms.

\paragraph{3. Regression Adjustment with Propensity Score}

As an additional robustness check, we estimated a parametric regression model that includes both the treatment indicator and the estimated propensity score as covariates \citep{austin2011introduction}. Specifically, the regression model is formulated as: 
\[
\log(\text{Salary}_i) = \beta_0 + \beta_1 * \text{Gender}_i + \beta_2 * \hat e_i + \epsilon_i,
\]
where \(\hat e_i\) denotes the estimated propensity score obtained from the logistic regression model.

All three approaches rely on the same propensity score specification but differ in their estimation strategies \citep{austin2011introduction}. As a result, they target different estimands depending on how observations are weighted or retained (e.g., matched samples vs. weighted populations) \citep{allan2020psmiptw}. More specifically, propensity score matching (PSM) targets the average treatment effect on the treated (ATT), as it focuses on treated units and their matched counterparts with similar covariate profiles. In contrast, inverse probability of treatment weighting (IPTW) and regression adjustment using the propensity score aim to estimate the average treatment effect (ATE) in the overall population by reweighting or modeling all observed units. These estimands differ in their target populations and interpretations. The ATT captures the effect of the treatment for individuals who actually received the treatment, and is therefore most relevant for evaluating the impact of treatment among the treated group. In contrast, the ATE represents the expected effect if the entire population were hypothetically assigned to treatment versus control, and is thus more appropriate for policy evaluation or population-level inference \citep{lu2025fourtargets}. In terms of implementation, matching emphasizes covariate balance by restricting attention to comparable units, IPTW utilizes the full sample but may be sensitive to extreme weights, and regression adjustment relies more heavily on model specification \citep{propensityScoreMatching2020}. Taken together, these methods provide complementary evidence and serve as robustness checks for the estimated gender effect.

\subsection{Non-Parametric Method: Causal Forest}

To model complex relationships and allow for treatment effect heterogeneity, we employed a non-parametric method known as the \textit{Causal Forest}, developed by \citet{athey2016recursive} and extended by \citet{wager2018estimation}. Unlike traditional regression-based approaches, which impose parametric assumptions and typically focus on population-level average treatment effects, causal forests estimate the \textit{Conditional Average Treatment Effect} (CATE), denoted as \( \tau(X_i) \), for each unit based on its observed covariates. A natural way to express this is as an ensemble average across trees:
\[
\hat{\tau}(X_i) = \frac{1}{T} \sum_{t=1}^{T} \hat{\tau}_t(X_i),
\]
where \( \hat{\tau}_t(X_i) \) is the treatment effect estimate from the \(t\)-th tree. This formulation emphasizes that the CATE is a point estimate conditional on covariates \(X_i\), obtained as a weighted average across trees.

In practice, however, causal forests are not simply tree averages. Following the generalized random forest framework \citep{athey2019estimating}, they are implemented as an orthogonalized local CATE estimator:
\[
\hat{\tau}(x) = \frac{\sum_{i=1}^n \alpha_i(x)\,\big(Y_i - \hat{m}^{(-i)}(X_i)\big)\,\big(Z_i - \hat{e}^{(-i)}(X_i)\big)}{\sum_{i=1}^n \alpha_i(x)\,\big(Z_i - \hat{e}^{(-i)}(X_i)\big)^2},
\]
where \( \alpha_i(x) \) are adaptive weights determined by the forest structure and can be interpreted as a data-driven kernel, \( \hat{m}^{(-i)}(X_i) \) is the estimated conditional mean outcome and learned from a separate subsample, and \( \hat{e}^{(-i)}(X_i) \) is the estimated propensity score. The superscript \((-i)\) indicates that the estimates are obtained without observation \(i\), which prevents overfitting. This orthogonalization, also known as double machine learning, step reduces overfitting and bias from nuisance parameter estimation. Moreover, this construction yields a doubly robust estimator, meaning that valid estimation of treatment effects is retained if either the outcome or the propensity nuisance function is correctly specified.

Beyond this estimator, causal forests incorporate several algorithmic innovations adjusted for causal inference. First, the algorithm adopts an honest estimation strategy through sample-splitting: one portion of the data is used to determine tree structure (i.e., splitting rules), and a separate, disjoint portion is used to estimate treatment effects within each leaf. This reduces overfitting and improves the validity of inference. Second, the method incorporates orthogonalization, also known as double machine learning, to reduce sensitivity to nuisance parameter estimation. In practice, this means that the model first estimates the propensity score and the conditional expectation of the outcome, then fits the forest using residualized versions of the treatment and outcome variables. Third, treatment effect estimation is localized—each tree partition defines neighborhoods of similar observations, and the treatment effect is estimated by comparing outcomes between treated and control units within these local regions. This enables the method to flexibly model high-order, nonlinear interactions without the need for manual specification.

In our analysis, the covariate matrix \( X_i \) included academic title, department code, university code, years of experience, and log-transformed i10-index. For the subset of faculty with valid Google Scholar IDs, we used this specification directly, without including the ID variable. For the full sample, however, we aligned the nuisance models with the parametric framework: the propensity nuisance \(\hat{e}(X)\) was estimated without Google Scholar ID, while the outcome nuisance \(\hat{m}(X,Z)\) incorporated it as a binary indicator. This ensured comparability across approaches and improved covariate balance, since including Google Scholar ID in the propensity score stage led to a less balanced post-matching.

We implemented causal forests using the \texttt{grf} package in R \citep{athey2019generalized}. For the subset sample, we grew 3,000 trees to stabilize estimates, used honest splitting to reduce bias, and set the minimum node size to 5 and the subsample fraction to 0.5, following common practice in causal forest estimation. We tuned only the number of covariates considered at each split ($mtry$), which balances predictive accuracy and computational efficiency. For the full sample, all major hyperparameters were tuned automatically to accommodate the higher-dimensional nuisance functions. To address extreme estimated propensities close to 0 or 1, we focused on the average treatment effect for the overlap population, which emphasizes comparisons among faculty with comparable treatment probabilities and yields more reliable inference than the full-sample ATE.

This method is particularly well-suited to our research question for several reasons. First, salary determination in academia is influenced by complex interactions between individual- and institution-level factors, such as rank, experience, publication record, and university type. These relationships are unlikely to be well-approximated by linear or additive models. The causal forest naturally accommodates such complexity in a data-driven, nonparametric way. Second, the gender effect on salary is unlikely to be homogeneous across faculty members. For instance, the disparity may differ between early-career and senior faculty, or between high-performing and less research-active scholars. The causal forest allows us to estimate individualized treatment effects and to explore how the effect of gender varies across different subpopulations, providing a nuanced perspective on inequality that average treatment effects may obscure.

\section{Results}\label{sec4}
 When analyzing the full data set, we included a binary indicator for Google Scholar ID in the outcome models. For individuals without matched Google Scholar IDs, a key confounder, the i10-index, is missing. To address this, we imputed missing values using the mean values calculated from similar background faculties with available information. This allows us to retain a larger sample size while still controlling for confounding factors in a consistent manner. As shown in Table~\ref{appendix:ols_regression}, the coefficient for this indicator is statistically insignificant ($p = 0.953$), suggesting that its presence does not independently predict salary. Therefore, in the rest of this section we focus on the subset of faculty with matched Google Scholar IDs, enabling more precise adjustment for research productivity. For completeness we present the results of the full data set in the Appendix~\ref{app:summary_full}.

\subsection{Non-Causal Analysis}
We first compared the raw (unadjusted) salaries between male and female faculty members in the subset with Google Scholar information. The average salary for male is \$134,169 and for female is \$118,460, which leads to a 11.71\% wage gap between two genders. The salary distribution is right-skewed, with more high-salary outliers among male faculty. Visualizations (see Appendix~\ref{appendix:salary_dist}) confirm the distributional differences.

\paragraph{OLS with main effects} 
The main effect Ordinary Least Squares (OLS) regression results, presented in Table~\ref{app:baseline_regression}, indicate that even after adjusting for these covariates, a statistically significant gender pay gap persists. The estimated coefficient for gender is \(-0.0321\) ($p = 2.32e^{-16}$), suggesting that female faculty members earn approximately 7.12\% less than male faculty members with similar observed characteristics.

\paragraph{OLS with interaction terms}

To allow for greater flexibility in how observed characteristics relate to salary, we extended the baseline specification by incorporating interaction terms between faculty rank and working years, as well as between academic performance, discipline, and institutional classification. This specification aligns with the structure used in the subsequent causal analyses, enabling direct comparison across methods, and captures heterogeneous returns across career stages, disciplines, and institutions. The comprehensive results are presented in Table~\ref{tab:ols_interaction_full}.\\The estimated gender coefficient remains negative and statistically significant ($\beta = -0.0289$, $p = 1.32e^{-11}$), indicating that female faculty earn approximately 6.44\% less than male faculty conditional on observed characteristics. This suggests that the gender pay gap is not fully explained by differences in observable characteristics. \\The interaction between faculty rank and working years reveals important differences in salary trajectories across career stages. \textit{Assistant Professors} exhibit limited or even declining salary growth over time ($\beta = -0.0087$, $p < 2e^{-16}$), consistent with early-career salary compression. \textit{Associate Professors} show modest improvements ($\beta = -0.0027$, $p < 2e^{-16}$), while \textit{Full Professors} experience positive and sustained salary growth ($\beta = 0.0025$, $p < 2e^{-16}$), reflecting the increasing returns to seniority and accumulated reputation. \\We also observed substantial heterogeneity in the returns to academic productivity across disciplines and institution types. In \textit{Business} and \textit{Medical and Health Sciences}, particularly at research-intensive (R1) institutions, higher publication impact is strongly associated with higher salaries (Business: $\beta = 0.1815$, $p < 2e^{-16}$ and Medical and Health Sciences: $\beta = 0.1418$, $p < 2e^{-16}$), reflecting the central role of research output in these fields. In contrast, in disciplines such as \textit{Technology and Engineering}, \textit{Social Sciences}, and \textit{Natural Sciences}, the returns to productivity depend critically on institutional context, with positive effects concentrated in research-oriented environments (Technology and Engineering: $\beta = 0.0404$, $p < 2e^{-16}$, Social Sciences: $\beta = 0.0388$, $p = 4.60e^{-11}$, and Natural Sciences: $\beta = 0.0376$, $p < 2e^{-16}$) and negative associations in teaching-focused institutions (in DRU(H) institutes, Technology and Engineering: $\beta = -0.0072$, $p = 0.236$, Social Sciences: $\beta = -0.0555$, $p < 2e^{-16}$, and Natural Sciences: $\beta = -0.0507$, $p = 9.33e^{-10}$; in bachelor/master institutes, Technology and Engineering: $\beta = -0.0567$, $p = 2.38e^{-7}$, Social Sciences: $\beta = -0.0698$, $p = 2.33e^{-9}$, and Natural Sciences: $\beta = -0.0646$, $p = 9.33e^{-10}$). In the \textit{Arts and Humanities}, productivity measured by citation-based metrics is consistently negatively (bachelor/master: $\beta = -0.0937$, $p < 2e^{-16}$, DRU(H): $\beta = -0.0776$, $p < 2e^{-16}$, and DU/VA: $\beta = -0.0484$, $p = 6.63e^{-11}$) associated with salary, suggesting that such metrics may not adequately capture scholarly impact in these fields. \\The model explains a substantial portion of the variation in salaries ($R^2 \approx 0.45$), indicating that the included covariates and interaction structure capture important determinants of faculty compensation. Overall, these results highlight that salary determination in academia is highly heterogeneous and context-dependent. Importantly, even after accounting for these complex interaction structures, a persistent gender pay gap remains, indicating that the observed disparity cannot be fully explained by differences in observable institutional, disciplinary, or career-related characteristics.

\subsection{Parametric Causal Analysis: Propensity Score-based Methods}

We first assessed the validity of key causal inference assumptions (unconfoundedness, positivity, and consistency). In particular, covariate balance diagnostics and regression specifications can be found in Appendix~\ref{appendix:assumptions}. 

\subsubsection{Average Treatment Effect on the Treated (ATT): Propensity Score Matching}

Using subset only including professors with Google Scholar information, we applied Propensity Score Matching (PSM) with logistic regression as the parametric causal inference method. Through an iterative model development process, we selected interaction terms $(\text{University Code} * \text{Department Code} * \log_{10}(\text{i10index}))$ and $(\text{Titles} * \text{Working Years})$. These interactions significantly improved model fit and balance, with the final model achieving the best explanatory power ($R^2 = 0.433$) and covariate balance (post-matching standardized differences near zero for all variables). Alternative models without these interactions showed poorer fit and less balanced matching, underscoring the importance of modeling complex relationships in academic salary structures.

We evaluated the matching process using numerical and visual methods, ensuring well-balanced treatment and control groups. The sample consists of 3,062 male and 1,920 female faculty members. After matching, 1,919 females were successfully paired with 1,179 comparable males, leaving 1,883 males and 1 female unmatched. The balance diagnostics confirm that post-matching standardized mean differences (SMD) are near zero for all key covariates, ensuring comparability between treatment and control groups. The full covariate balance table is available in Table~\ref{tab:full_balance}. As previously demonstrated, the Love Plot (Figure~\ref{fig:love_plot}) confirms that propensity score matching successfully balances key covariates across gender groups.

We then performed a linear regression on the matched data (see Table~\ref{app:psm_regression} for full results). Gender remains a significant predictor of salary, with a coefficient of $\beta = -0.0297$ (SE $= 0.0054$, $p = 3.61e^{-8}$), indicating that female faculty earn approximately 6.61\% less than their male counterparts with similar backgrounds. This confirms a persistent gender wage gap, even after accounting for institutional, departmental, and career-related factors, emphasizing the need for targeted interventions in academic salary structures. Our results also showed that career stage, research productivity, and disciplinary context jointly shape salary.

Overall, our PSM model yielded an $R^2 = 0.4368$, an adjusted $R^2 = 0.4328$, and a residual standard error of 0.1442 (df = 3075), indicating that the model explains 43.68\% of the variation in log-transformed salaries. The persistence of a significant gender wage gap (6.61\%, $p = 3.61e^{-8}$) even after controlling for multiple factors highlights the presence of structural inequities in academic compensation. 

\subsubsection{Average Treatment Effect (ATE): IPTW and Regression Adjustment}

To assess the robustness of our findings and examine whether the estimated gender effect generalizes beyond the matched sample, we additionally estimated the average treatment effect (ATE) using inverse probability of treatment weighting (IPTW) and regression adjustment based on the propensity score.

Using IPTW, which reweights the sample to approximate the overall population, we estimated a gender coefficient of $\beta = -0.0276$, corresponding to an approximate 6.17\% lower salary for female faculty relative to comparable male counterparts. Similarly, regression adjustment using the estimated propensity score yielded a comparable estimate of $\beta = -0.0290$, corresponding to a 6.45\% salary gap.

These estimates are highly consistent with the ATT obtained from propensity score matching, suggesting that the observed gender wage gap is not driven by sample restriction or lack of overlap, but instead reflects a persistent pattern across the broader population. Despite differences in estimation strategies, IPTW relying on weighting and regression adjustment relying on model specification, the results are qualitatively similar, reinforcing the robustness of our findings.

\subsection{Non-Parametric Causal Estimation: Causal Forest}

Using the Causal Forest model as a non-parametric causal inference method, we analyzed both overall trends and individual variations in gender-based salary disparities. The validity of these estimates depends on the assumption that, conditional on observed covariates, including rank, department, institutional affiliation, working years, and research productivity, gender assignment is unconfounded. The honesty principle embedded in causal forests helps reduce overfitting and bias in treatment effect estimation by separating model training from effect prediction \citep{wager2018estimation}.

For the subset of faculty with valid Google Scholar IDs, the estimated Average Treatment Effect (ATE) for the overlap population was $-0.0267$ (SE = $0.0039$), indicating that, among comparable faculty, female faculty earn  on average 5.96\% less than male faculty after adjusting for covariates. This confirms a persistent gender wage gap even after controlling for critical confounding variables. Sensitivity checks varying key hyperparameters yielded substantively unchanged results (see Appendix~\ref{app:cf_robustness}).

To explore individual-level variation, we estimated Individual Treatment Effects (ITEs) using the causal forest model. The distribution of ITEs, shown in Appendix~\ref{app:ite_dist}, ranges from $-0.035$ to $-0.021$, underscoring that the gender effect on salaries is far from uniform. Such heterogeneity indicates that average estimates may conceal important subgroup differences, motivating a closer examination of where disparities are most obvious.

We visualized ITEs by coloring observations with different covariates to detect systematic patterns. Among the explored factors, department and research productivity revealed the clearest structure, with Medical and Health Sciences (MHS) standing out as the field where gender inequities are particularly acute.

\begin{figure}[H]
    \centering
    \includegraphics[width=0.75\textwidth]{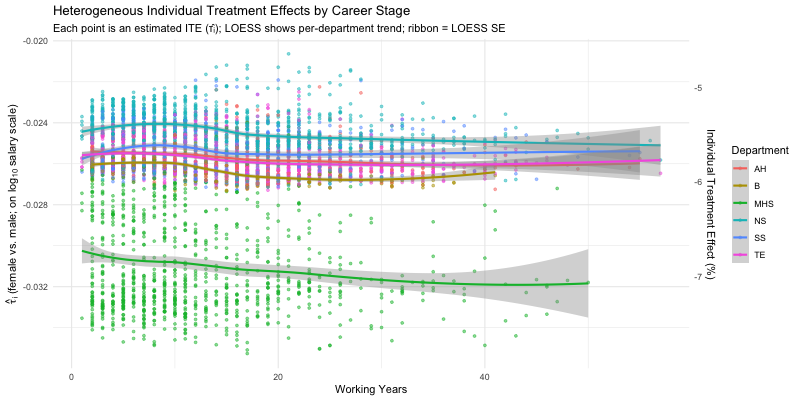}
    \caption{Heterogeneous Treatment Effects by Working Years.}
    \label{tab:working_years}
\end{figure}

Figure~\ref{tab:working_years} displays ITEs by career stage across departments. In most fields, the gender gap remains relatively stable over the career path, but MHS faculty show a consistently larger salary disadvantage for women. On average, women in MHS earn about 7\% less than comparable men, compared to least gaps around 5.5\% in Natural and Social Sciences (NS, SS) and about 6\% in Arts and Humanities (AH), Business (B), and Technology and Engineering (TE). The separation between MHS and other disciplines widens with experience, suggesting persistent structural barriers in this field.

\begin{figure}[H]
    \centering
    \includegraphics[width=0.75\textwidth]{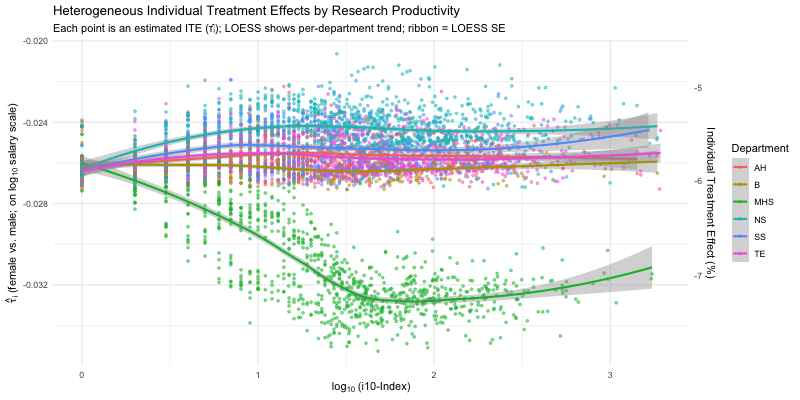}
    \caption{Treatment Effects by Research Productivity ($\log_{10}$ i10-index).}
    \label{tab:research_impact}
\end{figure}

Figure~\ref{tab:research_impact} conditions on research productivity. For MHS faculty, the estimated penalty starts around 6\% at lower productivity levels, worsens to more than 7\% in the mid-range, and then shows slight improvement at the highest levels, though still remaining near 7\%. By contrast, other departments display relatively little variation, with gaps largely concentrated between 5--6\%. This pattern suggests that women in MHS face a particularly persistent and pronounced disadvantage, especially at intermediate levels of research productivity, whereas disparities in other fields are smaller and more stable.

These findings reinforce the view that structural disparities in faculty compensation are not merely average effects but reflect deeper patterns shaped by tenure, institutional norms, and how scholarly output is translated into financial reward. Non-parametric causal estimation thus provides a flexible tool for identifying when and where inequities are most acute, offering valuable insights for both empirical researchers and academic administrators.

\subsection{Comparison Across Methods}
Table \ref{tab:summary_table} summarizes the estimated gender pay gap across multiple methods. Overall, the results are highly consistent: adjusted estimates of the gender gap range from approximately 6\% to 7\%, substantially smaller than the unadjusted gap of 11.71\%, indicating that a large portion of the raw disparity can be explained by observed covariates.

Despite this overall consistency, two notable patterns emerge across methods. First, the estimate obtained via propensity score matching (PSM) is slightly larger than those from regression-based and weighting approaches. This difference is expected, as PSM targets the average treatment effect on the treated (ATT), whereas regression adjustment and inverse probability weighting (IPTW) target the average treatment effect (ATE). In this context, the ATT corresponds to the gender effect for female faculty, and the larger estimate suggests that the pay gap may be more pronounced within this subgroup compared to the overall population.

Second, the causal forest estimate is consistently smaller than those from parametric methods. This may reflect the ability of the causal forest to capture treatment effect heterogeneity and nonlinear relationships between covariates and outcomes. It also highlights the importance of allowing for flexible, nonparametric structures when estimating causal effects in complex observational settings.

Taken together, the close agreement across methods provides reassuring evidence of robustness, while the observed differences offer additional insight into the role of estimand choice and model flexibility in shaping causal conclusions.

\begin{table}[H]
    \centering
    \caption{Estimated Gender Pay Gaps ($\log_{10}$ salary scale) by Method for Google Scholar Subset Data. Estimates are reported with 95\% confidence intervals.}
    \label{tab:summary_table}
    \begin{tabular}{llcc}
    \toprule
    \textbf{Method} & \textbf{Estimate $\beta$ (95\% CI)} & \textbf{Estimated Gap (\%)} \\
    \midrule
    Unadjusted Analysis & \textemdash{} & 11.71\% \\
    Baseline OLS Regression & -0.0321 (-0.0398, -0.0245) & 7.12\% (5.49\%, 8.76\%)\\
    OLS with Interaction Terms & -0.0289 (-0.0372, -0.0205) & 6.44\% (4.61\%, 8.21\%)\\
    PSM & -0.0297 (-0.0402, -0.0191) & 6.61\% (4.30\%, 8.84\%) \\
    IPTW & -0.0276 (-0.0355, -0.0198) & 6.17\% (4.46\%, 7.85\%)\\
    Regression Adjustment & -0.0290 (-0.0400, -0.0179) & 6.45\% (4.04\%, 8.80\%)\\
    Causal Forest & -0.0267 (-0.0343, -0.0191) & 5.96\% (4.30\%, 7.59\%)\\
    \bottomrule
    \end{tabular}
\end{table}

\section{Conclusion and Discussion}\label{sec5}

Descriptive comparisons reveal that, on average, female faculty earn 11.71\% less than male colleagues. Even after adjusting for institutional, disciplinary, and productivity-related factors using causal inference methods, substantial gaps remain: propensity score matching yields a disparity of 6.61\%, inverse probability of treatment weighting produces a gender gap of 6.17\%, regression adjustment with propensity score indicates a 6.45\% difference, and causal forests estimate an average treatment effect of 5.96\%. These results provide obvious evidence of structural gender-based salary inequities across the North Carolina public university system.

The gender wage gap is not uniform. Heterogeneity analyses show that disparities vary systematically across departments, with Medical and Health Sciences (MHS) standing out as the field where inequities are most acute. Female faculty in MHS earn approximately 7\% less than comparable men, a penalty that persists across career stages and remains visible even when conditioning on research productivity. By contrast, gaps in Natural Sciences and Social Sciences are closer to 5.5\%, while Arts and Humanities, Business, and Technology and Engineering average around 6\%. This suggests that women in MHS face more persistent structural barriers to pay equity, whereas disparities in other fields, while present, are less severe.

Institutional and disciplinary contexts also shape compensation patterns. Business and Medical and Health Sciences departments, particularly at R1 (DU/VA) institutions, demonstrate the strongest positive returns to research productivity. In contrast, the Arts and Humanities exhibit consistently negative associations between salary and citation-based metrics, raising concerns about the suitability of bibliometric indicators in evaluating scholarly impact in these fields. Other fields (Technology and Engineering, Natural Sciences, Social Sciences) display mixed patterns: research output is rewarded at R1 institutions but often penalized or ignored at less research-intensive settings. These findings underscore the interaction between institution type, academic discipline, and evaluative metrics in determining pay structures.

Our findings largely corroborate prior work documenting persistent gender pay gaps in academia, including evidence that adjusted disparities typically remain in the single-digit range even after accounting for rank, experience, and productivity \citep{kim2024gendergap}. At the same time, this study contributes new evidence along two key dimensions. First, by combining propensity score–based parametric methods with causal forests on a comprehensive faculty census from a single statewide public system, we move beyond descriptive or regression-based analyses and provide more credible causal estimates of the gender penalty, along with a detailed characterization of its heterogeneity across disciplines and individual characteristics. In this sense, while the overall magnitude of the gap is consistent with prior studies \citep{chen2019gender}, our approach strengthens identification and reveals substantial heterogeneity in the gender wage gap across disciplines and productivity levels, which would be obscured by focusing only on average treatment effects. Second, our results uncover systematic heterogeneity in how the gender wage gap varies across academic contexts. Consistent with existing research in academic medicine \citep{catenaccio2022addressing, miller2022gender}, we find that Medical and Health Sciences exhibit the most pronounced disparities, while other fields show smaller but still persistent gaps. Beyond this, we show that the magnitude of the gender penalty varies with research productivity in a discipline-specific manner: in Medical and Health Sciences, the gap tends to widen at intermediate levels of productivity, whereas in other fields it remains relatively stable. Taken together, these findings suggest that the gender wage gap in academia is not only persistent but also structurally heterogeneous, shaped by the interaction between disciplinary norms, career paths, and how scholarly output is evaluated within different fields.

Taken together, our findings highlight the need for comprehensive policy reforms. Institutions should improve salary transparency, standardize criteria for evaluating research output across disciplines, and conduct regular equity audits disaggregated by career stage, institution type, and academic field. Targeted interventions, such as early-career mentoring, off-scale salary review processes, and promotion pathway monitoring, may help address disparities before they become entrenched.

This study has several limitations. First, academic productivity is measured solely through Google Scholar, which may underrepresent contributions such as teaching excellence, mentorship, and institutional service \citep{ACE2022equity}. Second, the analysis is cross-sectional and does not capture how disparities evolve over time. Third, while the UNC system provides a coherent institutional framework, the findings may not generalize to private universities or public systems with different governance structures \citep{kuhn2019feature}. Future work would benefit from longitudinal analyses, broader definitions of scholarly impact, and investigation of intersectional factors, such as race, marital status, and contract type, that may further shape pay equity in academia.

\section*{Code availability}
All code used for data cleaning, analysis, and visualization in this study is publicly available on GitHub at:
\newline
\url{https://github.com/tinazhangzh/gender-pay-gap}. The repository includes Python scripts for data extraction and preprocessing; R scripts for statistical modeling, matching, and causal forest analysis; and supplementary materials including balance diagnostics and subgroup plots. 

\bibliography{References}
\bibliographystyle{plainnat} 

\appendix
\section{Variable Descriptions}
\subsection{Distribution of Gender Assignment Methods}
Table~\ref{tab:gender_distribution} presents the distribution of gender assignment methods employed in this study. The majority of cases were automatically assigned with high confidence, while a small subset required manual verification due to low confidence scores, undetermined gender, or uncommon/culturally specific names.
\begin{table}[H]
    \centering
    \renewcommand{\arraystretch}{1.2}
    \begin{tabular}{p{7cm}cc}
        \hline
        \textbf{Assignment Method} & \textbf{Number of Cases} & \textbf{Percentage (\%)} \\
        \hline
        Automated Assignment ($\geq$60\% confidence) & 11505 & 95.56 \\
        \hline
        \multicolumn{3}{l}{\textbf{Manual Verification:}} \\
        \hspace{5mm} Names with confidence $<$60\% & 274 & 2.28 \\
        \hspace{5mm} Undetermined gender cases & 164 & 1.36 \\
        \hspace{5mm} Uncommon/culturally specific names & 96 & 0.8 \\
        \hline
        \textbf{Total} & \textbf{12039} & \textbf{100} \\
        \hline
    \end{tabular}
    \caption{Distribution of Gender Assignment Methods}
    \label{tab:gender_distribution}
\end{table}

\subsection{Supplementary Table}
Table~\ref{tab:summary_nc_professors} provides descriptive information on the 16 universities in the University of North Carolina system, including faculty counts, Carnegie classifications, and average salary levels as of 2022.
\begin{table}[H]
\centering
\small 
\resizebox{\textwidth}{!}{
\begin{tabular}{p{4.8cm} r l r r}
\hline
\textbf{University Name} & \textbf{Number of} & \textbf{Carnegie} & \textbf{Mean} & \textbf{Median} \\
                         & \textbf{Professors} & \textbf{Classification} & \textbf{Salary} & \textbf{Salary} \\
\hline
\raggedright Appalachian State University & 760 & DRU(H) & \$87,467 & \$81,880 \\
\raggedright Elizabeth City State University & 77 & master & \$76,156 & \$73,230 \\
\raggedright East Carolina University & 1394 & DRU(H) & \$132,913 & \$94,201 \\
\raggedright Fayetteville State University & 206 & master & \$85,044 & \$77,493 \\
\raggedright NC Agricultural and Technical State University & 378 & DRU(H) & \$92,410 & \$85,425 \\
\raggedright NC Central University & 277 & master & \$90,888 & \$83,000 \\
\raggedright NC State University & 1783 & DU/VA & \$122,989 & \$118,566 \\
\raggedright UNC Asheville & 139 & bachelor & \$83,648 & \$83,214 \\
\raggedright UNC Chapel Hill & 3919 & DU/VA & \$175,131 & \$152,250 \\
\raggedright UNC Charlotte & 875 & DRU(H) & \$106,446 & \$95,919 \\
\raggedright UNC Greensboro & 655 & DRU(H) & \$91,681 & \$82,439 \\
\raggedright UNC Pembroke & 220 & master & \$79,245 & \$73,041 \\
\raggedright UNC School of the Arts & 117 & bachelor
 & \$77,004 & \$74,928 \\
\raggedright UNC Wilmington & 593 & DRU(H) & \$89,145 & \$81,039 \\
\raggedright Western Carolina University & 443 & master & \$84,746 & \$77,327 \\
\raggedright Winston-Salem State University & 203 & bachelor & \$84,622 & \$78,864 \\
\hline
\end{tabular}
}
\caption{Institutions in the University of North Carolina System as of 2022}
\label{tab:summary_nc_professors}
\end{table}

\section{Assumption Validation for Causal Inference}
\label{appendix:assumptions}

To ensure the validity of our causal inference analysis, we assessed the three standard assumptions: unconfoundedness, positivity, and consistency. This appendix summarizes the diagnostics and model checks used to support these assumptions.

\subsection{Unconfoundedness: Covariate Balance Before and After Matching}

The unconfoundedness assumption requires that there are no unmeasured confounders affecting both treatment (gender) and outcome (salary), conditional on observed covariates. To approximate this assumption, we applied Propensity Score Matching (PSM) and examined covariate balance between male and female faculty before and after matching.

Table~\ref{tab:full_balance} presents the standardized mean differences (SMD) for each covariate. The matching process substantially reduced imbalance across academic rank, experience, institutional affiliation, and research productivity. All post-matching SMD values fall below 0.1, indicating acceptable balance.

\begin{table}[H]
    \centering
    \renewcommand{\arraystretch}{1.1}
    \begin{tabular}{lcc}
        \hline
        \textbf{Covariate} & \textbf{SMD Before Matching} & \textbf{SMD After Matching} \\
        \hline
        Titles (Assistant Professor) & 0.1239 & 0.0052 \\
        Titles (Associate Professor) & 0.0238 & -0.0162 \\
        Titles (Professor) & -0.1477 & 0.0109 \\
        Working Years & -0.3181 & -0.0100 \\
        University Code (bachelor/master) & -0.0128 & 0.0052 \\
        University Code (DRU(H)) & 0.0127 & -0.0198 \\
        University Code (DU/VA) & 0.0001 & 0.0146 \\
        Department Code (AH) & 0.0289 & 0.0036 \\
        Department Code (B) & -0.0188 & -0.0047 \\
        Department Code (MHS) & 0.0815 & 0.0068 \\
        Department Code (NS) & -0.0431 & -0.0120 \\
        Department Code (SS) & 0.0468 & 0.0052 \\
        Department Code (TE) & -0.0954 & 0.0010 \\
        log$_{10}$(i10-index) & -0.4274 & 0.0062 \\
        \hline
    \end{tabular}
    \caption{Covariate balance before and after matching. Standardized mean differences (SMD) below 0.1 indicate acceptable balance.}
    \label{tab:full_balance}
\end{table}

\begin{figure}[H]
  \centering
  \includegraphics[width=0.75\textwidth]{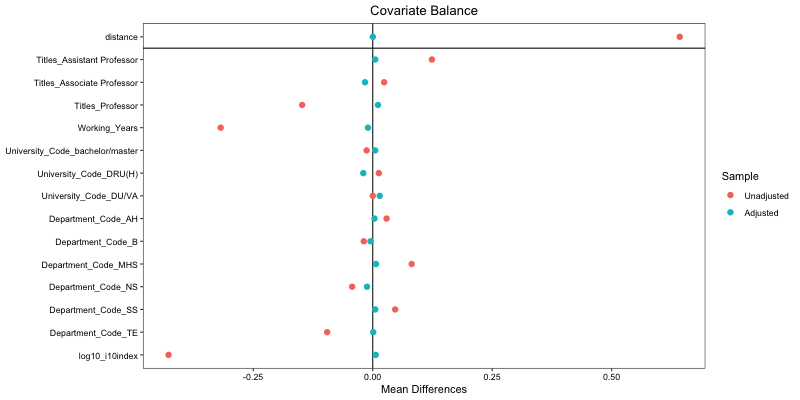}
  \caption{Love Plot illustrating standardized mean differences (SMD) before and after matching. Values closer to zero indicate improved covariate balance post-matching.}
  \label{fig:love_plot}
\end{figure}

To visualize the improvement in balance, Figure~\ref{fig:love_plot} presents a Love plot showing SMDs before and after matching.

\begin{figure}[H]
  \centering
  \includegraphics[width=0.75\textwidth]{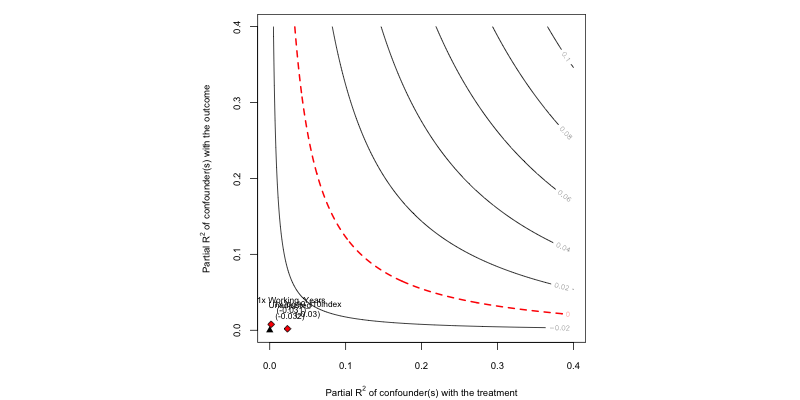}
  \caption{Sensitivity analysis to unobserved confounding. Contours show combinations of partial $R^2$ of a hypothetical confounder with the treatment and outcome that yield the same adjusted estimate; the red dashed curve indicates where the estimated effect is reduced to zero.}
  \label{fig:ovb}
\end{figure}

We conducted a sensitivity analysis to assess the potential impact of unobserved confounding using the robustness value framework. The estimated treatment effect of gender on log salary was $-0.032$ (SE = 0.0039). The robustness value (RV) for the point estimate is 0.110, indicating that an unobserved confounder would need to explain at least 11.0\% of the residual variance in both treatment assignment and the outcome to fully account for the observed effect. At the 5\% significance level, the robustness value is 0.085, meaning that a confounder explaining more than 8.5\% of residual variance in both treatment and outcome would be required to render the effect statistically insignificant. For comparison, in figure~\ref{fig:ovb}, observed covariates such as working years and research productivity (log i10-index) exhibit substantially smaller partial $R^2$ values. This suggests that an unobserved confounder strong enough to overturn the results would need to be considerably more influential than the measured covariates, providing evidence that our findings are reasonably robust to hidden bias.

\subsection{Positivity: Propensity Score Overlap}

\begin{figure}[H]
    \centering
    \includegraphics[width=0.75\textwidth]{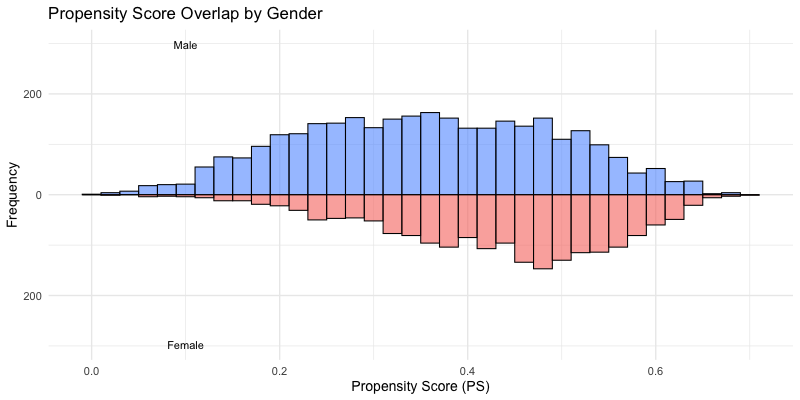}
    \caption{Propensity Score Overlap by Gender}
    \label{fig:ps_overlap}
\end{figure}

The positivity assumption requires that each unit has a non-zero probability of receiving either treatment. We assessed this assumption by plotting the distribution of estimated propensity scores across treatment groups. Figure~\ref{fig:ps_overlap} shows substantial overlap between male and female faculty, indicating that the positivity assumption holds.

\subsection{Consistency: Treatment Definition and No Interference}

The consistency assumption requires that the treatment is well-defined and that there is no interference between units. In our study, the binary variable \texttt{Gender} is clearly defined as 0 (Male) or 1 (Female) for all observations, ensuring that treatment status is consistent and unambiguous.

To address potential violations of the no-interference assumption, especially institutional spillover effects, we included university and department fixed effects in the outcome regression model:

\begin{equation*}
    \log(\text{Salary}) = \beta_0 + \beta_1 \text{Gender} + \sum_j \gamma_j \text{University}_j + \sum_k \delta_k \text{Department}_k + X\beta + \varepsilon
\end{equation*}

The results of this specification are presented in Table~\ref{tab:fe_results}, showing that gender remains statistically significant after controlling for institutional heterogeneity.

\begin{table}[H]
    \centering
    \begin{tabular}{lccc}
        \hline
        \textbf{Variable} & \textbf{Estimate} & \textbf{Std. Error} & \textbf{p-value} \\
        \hline
        Gender (Female) & -0.0321 & 0.0039 & 2.32e-16 *** \\
        University Fixed Effects & Yes & - & - \\
        Department Fixed Effects & Yes & - & - \\
        log$_{10}$(i10-index) & 0.0094 & 0.0031 & 0.00256 ** \\
        Titles (Associate Professor) & 0.0526 & 0.0049 & $<$2e-16 *** \\
        Titles (Professor) & 0.1760 & 0.0057 & $<$2e-16 *** \\
        Working Years & -0.0015 & 0.0002 & 6.31e-10 *** \\
        \hline
    \end{tabular}
    \caption{Fixed Effects Regression Controlling for Institutional and Department-Level Factors}
    \label{tab:fe_results}
\end{table}

While these fixed effects help account for unobserved institutional and departmental factors that may induce correlated outcomes across faculty, they do not fully eliminate the possibility of interference. In particular, mechanisms such as peer effects, mentorship structures, or gender composition within departments may influence hiring, promotion, and salary-setting processes. Therefore, the no-interference assumption should be interpreted as an approximation in this context.

Our estimates should thus be understood as capturing average differences between male and female faculty conditional on observed and institutional characteristics, rather than strictly isolating causal effects under complete independence across units.

Taken together, the assumptions of unconfoundedness, positivity, and consistency are plausibly satisfied in our setting, though they should be interpreted with appropriate caution given the potential for residual interference.

\section{Subset Analysis Results}
\label{app:sub_analysis}
Here, we present additional details on the analysis of subset of the data restricted to faculty with matched. Google Scholar IDs.

\subsection{Descriptive Statistics of Salary Distribution}
\label{appendix:salary_dist}
To supplement our raw wage difference analysis, we presented additional visualizations illustrating salary distribution patterns.

\begin{figure}[H]
   \centering
   \includegraphics[width=0.75\textwidth]{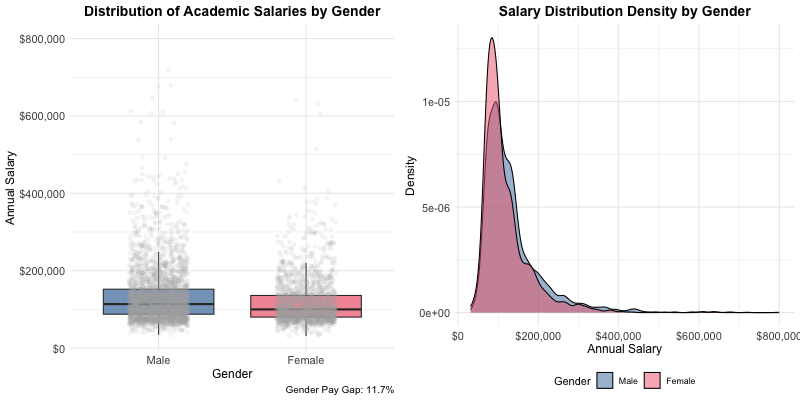}
   \caption{Distribution of Academic Salaries by Gender. Left panel shows a boxplot with individual observations (jittered), while the right panel displays the density distribution of salaries by gender.}
   \label{fig:salary_dist}
\end{figure}

The boxplot and density plots (Figure~\ref{fig:salary_dist}) highlight notable gender differences. The median salary for male faculty members is higher than that of female faculty, with a raw gender pay gap of 11.71\%. The density plot further reveals a right-skewed distribution, indicating the presence of high-salary outliers in both groups.

\begin{figure}[H]
   \centering
   \includegraphics[width=0.5\textwidth]{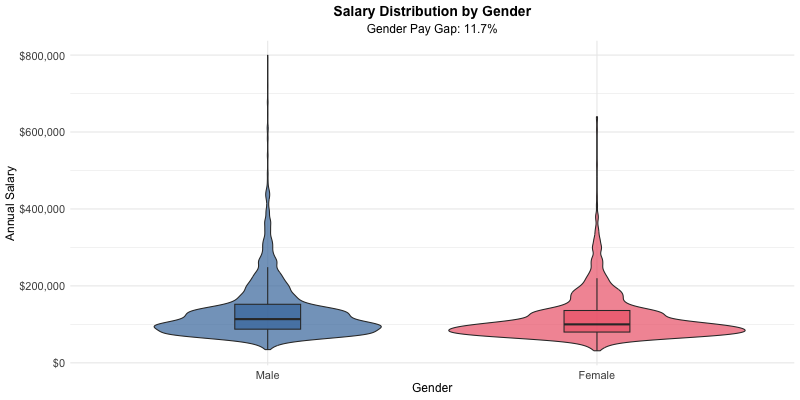}
   \caption{Violin plot illustrating the distribution of academic salaries by gender, with embedded boxplots displaying quartiles and median.}
   \label{fig:salary_violin}
\end{figure}

The violin plot (Figure~\ref{fig:salary_violin}) shows that male faculty salaries display greater variation, particularly in the upper range. The core salary range is similar between genders, but the overall distribution differs, with a more noticeable thinning in the female salary distribution at higher levels.

\subsection{Non-Causal Analysis}
\label{app:baseline_regression}
Table~\ref{tab:baseline_regression} reports the baseline regression results using log-transformed salary as the dependent variable, providing descriptive associations between gender, academic characteristics, and salaries prior to applying causal inference methods.
\begin{table}[H]
    \centering
    \renewcommand{\arraystretch}{1.2}
    \begin{tabular}{lcccc}
        \hline
        \textbf{Variable} & \textbf{Estimate} & \textbf{Std. Error} & \textbf{t-value} & \textbf{p-value} \\
        \hline
        (Intercept) & 4.7647 & 0.0088 & 539.751 & $<$2e-16*** \\
        Gender (Female) & -0.0321 & 0.0039 & -8.232 & 2.32e-16*** \\
        Titles: Associate Professor & 0.0526 & 0.0049 & 10.750 & $<$2e-16*** \\
        Titles: Professor & 0.1760 & 0.0057 & 30.753 & $<$2e-16*** \\
        University Code (DRU(H)) & 0.0567 & 0.0067 & 8.511 & $<$2e-16*** \\
        University Code (DU/VA) & 0.1693 & 0.0066 & 25.501 & $<$2e-16*** \\
        Department Code (B) & 0.2743 & 0.0083 & 33.198 & $<$2e-16*** \\
        Department Code (MHS) & 0.2674 & 0.0067 & 39.647 & $<$2e-16*** \\
        Department Code (NS) & 0.0754 & 0.0066 & 11.404 & $<$2e-16*** \\
        Department Code (SS) & 0.0748 & 0.0070 & 10.727 & $<$2e-16*** \\
        Department Code (TE) & 0.0971 & 0.0066 & 14.714 & $<$2e-16*** \\
        Working Years & -0.0015 & 0.0002 & -6.195 & 6.31e-10*** \\
        log$_{10}$(i10-index) & 0.0094 & 0.0031 & 3.018 & 0.00256** \\
        \hline
    \end{tabular}
    \caption{Baseline Regression Results: Log-Transformed Salary as Dependent Variable}
    \label{tab:baseline_regression}
\end{table}

\begin{table}[H]
\centering
\renewcommand{\arraystretch}{1.2}
\begin{tabular}{lcccc}
\hline
\textbf{Variable} & \textbf{Estimate} & \textbf{Std. Error} & \textbf{t-value} & \textbf{p-value} \\
\hline

(Intercept) & 5.0440 & 0.0064 & 785.655 & $<$2e-16*** \\

Gender (Female) & -0.0289 & 0.0043 & -6.783 & 1.32e-11*** \\

Assistant Professor $\times$ Years & -0.0087 & 0.0006 & -15.470 & $<$2e-16*** \\
Associate Professor $\times$ Years & -0.0027 & 0.0003 & -8.263 & $<$2e-16*** \\
Professor $\times$ Years & 0.0025 & 0.0002 & 11.196 & $<$2e-16*** \\

Bachelor/Master $\times$ AH $\times$ i10 & -0.0937 & 0.0106 & -8.848 & $<$2e-16*** \\
DRU(H) $\times$ AH $\times$ i10 & -0.0776 & 0.0066 & -11.740 & $<$2e-16*** \\
DU/VA $\times$ AH $\times$ i10 & -0.0484 & 0.0074 & -6.543 & 6.63e-11*** \\

Bachelor/Master $\times$ Business $\times$ i10 & 0.0140 & 0.0127 & 1.101 & 0.271 \\
DRU(H) $\times$ Business $\times$ i10 & 0.0817 & 0.0076 & 10.815 & $<$2e-16*** \\
DU/VA $\times$ Business $\times$ i10 & 0.1815 & 0.0090 & 20.083 & $<$2e-16*** \\

Bachelor/Master $\times$ MHS $\times$ i10 & -0.0141 & 0.0213 & -0.661 & 0.509 \\
DRU(H) $\times$ MHS $\times$ i10 & 0.0667 & 0.0067 & 9.883 & $<$2e-16*** \\
DU/VA $\times$ MHS $\times$ i10 & 0.1418 & 0.0044 & 32.253 & $<$2e-16*** \\

Bachelor/Master $\times$ NS $\times$ i10 & -0.0646 & 0.0105 & -6.132 & 9.33e-10*** \\
DRU(H) $\times$ NS $\times$ i10 & -0.0507 & 0.0062 & -8.207 & 2.87e-16*** \\
DU/VA $\times$ NS $\times$ i10 & 0.0376 & 0.0043 & 8.790 & $<$2e-16*** \\

Bachelor/Master $\times$ SS $\times$ i10 & -0.0698 & 0.0117 & -5.984 & 2.33e-09*** \\
DRU(H) $\times$ SS $\times$ i10 & -0.0555 & 0.0067 & -8.335 & $<$2e-16*** \\
DU/VA $\times$ SS $\times$ i10 & 0.0388 & 0.0059 & 6.598 & 4.60e-11*** \\

Bachelor/Master $\times$ TE $\times$ i10 & -0.0568 & 0.0110 & -5.174 & 2.38e-07*** \\
DRU(H) $\times$ TE $\times$ i10 & -0.0072 & 0.0060 & -1.185 & 0.236 \\
DU/VA $\times$ TE $\times$ i10 & 0.0404 & 0.0042 & 9.623 & $<$2e-16*** \\
\hline
\end{tabular}
\caption{OLS Regression with Interaction Terms}
\label{tab:ols_interaction_full}
\end{table}

\subsection{Parametric Causal Analysis}
\label{app:psm_regression}
Table~\ref{tab:psm_regression} presents the regression estimates from the parametric causal analysis with interaction terms, providing detailed coefficient results.
\begin{table}[H]
    \centering
    \renewcommand{\arraystretch}{1.2}
    \begin{tabular}{lcccc}
        \hline
        \textbf{Variable} & \textbf{Estimate} & \textbf{Std. Error} & \textbf{t-value} & \textbf{p-value} \\
        \hline
        (Intercept) & 5.0390 & 0.0083 & 604.473 & $<$2e-16*** \\
        Gender (Female) & -0.0297 & 0.0054 & -5.523 & 3.61e-08*** \\
        Titles: Assistant Professor $\times$ Working Years & -0.0083 & 0.0007 & -12.075 & $<$2e-16*** \\
        Titles: Associate Professor $\times$ Working Years & -0.0024 & 0.0004 & -5.651 & 1.74e-08*** \\
        Titles: Professor $\times$ Working Years & 0.0030 & 0.0003 & 9.215 & $<$2e-16*** \\
        bachelor/master $\times$ Dept. AH $\times$ log$_{10}$(i10-index) & -0.0873 & 0.0135 & -6.461 & 1.20e-10*** \\
        DRU(H) $\times$ Dept. AH $\times$ log$_{10}$(i10-index) & -0.0828 & 0.0086 & -9.671 & $<$2e-16*** \\
        DU/VA $\times$ Dept. AH $\times$ log$_{10}$(i10-index) & -0.0559 & 0.0109 & -5.155 & 2.69e-07*** \\
        bachelor/master $\times$ Dept. B $\times$ log$_{10}$(i10-index) & 0.0179 & 0.0210 & 0.852 & 0.3942 \\
        DRU(H) $\times$ Dept. B $\times$ log$_{10}$(i10-index) & 0.0938 & 0.0113 & 8.314 & $<$2e-16*** \\
        DU/VA $\times$ Dept. B $\times$ log$_{10}$(i10-index) & 0.1610 & 0.0121 & 13.323 & $<$2e-16*** \\
        bachelor/master $\times$ Dept. MHS $\times$ log$_{10}$(i10-index) & -0.0182 & 0.0282 & -0.646 & 0.5185 \\
        DRU(H) $\times$ Dept. MHS $\times$ log$_{10}$(i10-index) & 0.0646 & 0.0083 & 7.833 & 6.52e-15*** \\
        DU/VA $\times$ Dept. MHS $\times$ log$_{10}$(i10-index) & 0.1479 & 0.0056 & 26.450 & $<$2e-16*** \\
        bachelor/master $\times$ Dept. NS $\times$ log$_{10}$(i10-index) & -0.0764 & 0.0175 & -4.359 & 1.35e-05*** \\
        DRU(H) $\times$ Dept. NS $\times$ log$_{10}$(i10-index) & -0.0482 & 0.0089 & -5.430 & 6.08e-08*** \\
        DU/VA $\times$ Dept. NS $\times$ log$_{10}$(i10-index) & 0.0356 & 0.0061 & 5.855 & 5.28e-09*** \\
        bachelor/master $\times$ Dept. SS $\times$ log$_{10}$(i10-index) & -0.0758 & 0.0172 & -4.398 & 1.13e-05*** \\
        DRU(H) $\times$ Dept. SS $\times$ log$_{10}$(i10-index) & -0.0565 & 0.0087 & -6.462 & 1.20e-10*** \\
        DU/VA $\times$ Dept. SS $\times$ log$_{10}$(i10-index) & 0.0429 & 0.0076 & 5.636 & 1.89e-08*** \\
        bachelor/master $\times$ Dept. TE $\times$ log$_{10}$(i10-index) & -0.0509 & 0.0246 & -2.072 & 0.0384* \\
        DRU(H) $\times$ Dept. TE $\times$ log$_{10}$(i10-index) & -0.0769 & 0.0095 & -0.008 & 0.9935 \\
        DU/VA $\times$ Dept. TE $\times$ log$_{10}$(i10-index) & 0.0397 & 0.0061 & 6.520 & 8.19e-11*** \\
        \hline
    \end{tabular}
    \caption{Three-way Interaction Model: Gender, Academic Context, and Productivity (log$_{10}$(i10-index))}
    \label{tab:psm_regression}
\end{table}

\subsection{Non-Parametric Causal Analysis}
\subsubsection{Robustness Check for Different Hyperparameters}
\label{app:cf_robustness}
To assess robustness to hyperparameter choices, we re-estimated the causal forest model under alternative specifications of minimum node size, subsample fraction, number of covariates ($mtry$), and number of trees. As shown in Table~\ref{tab:cf_robustness}, the estimated Average Treatment Effect (ATE) remained stable across specifications, with values consistently around $-0.026$ and standard errors near 0.004. This indicates that our findings are not sensitive to tuning parameters and are robust to reasonable variations in model configuration.

\begin{table}[H]
    \centering
    \begin{tabular}{lcc}
        \toprule
        Specification & ATE Estimate & Standard Error \\
        \midrule
        Baseline (tuned $mtry$, sample.fraction = 0.49) & -0.0264 & 0.0039 \\
        Min node size = 10 & -0.0264 & 0.0039 \\
        Min node size = 20 & -0.0264 & 0.0039 \\
        Sample fraction = 0.45 & -0.0263 & 0.0039 \\
        $mtry = \lceil p/3 \rceil = 8$ & -0.0270 & 0.0040 \\
        $mtry = \lceil p/2 \rceil = 11$ & -0.0266 & 0.0039 \\
        Number of trees = 1500 & -0.0265 & 0.0039 \\
        Number of trees = 5000 & -0.0264 & 0.0039 \\
        \bottomrule
    \end{tabular}
    \caption{Robustness of Causal Forest ATE Estimates to Alternative Hyperparameter Specifications}
    \label{tab:cf_robustness}
\end{table}

\subsubsection{Distribution of Estimated Individual Treatment Effects}
\label{app:ite_dist}
\begin{figure}[H]
    \centering
    \includegraphics[width=0.8\textwidth]{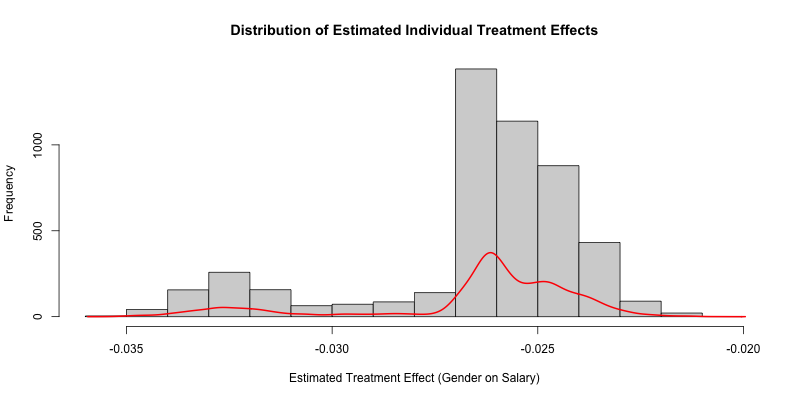}
    \caption{Distribution of Individual Treatment Effects}
    \label{fig:ite_dist}
\end{figure}

Figure~\ref{fig:ite_dist} shows the distribution of estimated individual treatment effects, illustrating the variation in salary impacts across faculty members.

\section{Comprehensive Analysis Results}
\label{app:summary_full}
This section reports detailed result for the entire dataset. We imputed the missing Google Scholar data with the mean values of faculty with similar characteristics, e.g. gender, department type and rank. This section shows similar results to the subset data analysis in Appendix~\ref{app:sub_analysis}.

\subsection{Non-Causal Analysis}
The full dataset yields similar patterns with the subset data analysis including only matched Google Scholar information: an unadjusted gender pay gap of 12.80\%, and comparable distribution shapes. Detailed plots for the full dataset are provided below in Figure~\ref{appendix:salary_dist2}.

\label{appendix:salary_dist2}
\begin{figure}[H]
   \centering
   \includegraphics[width=0.67\textwidth]{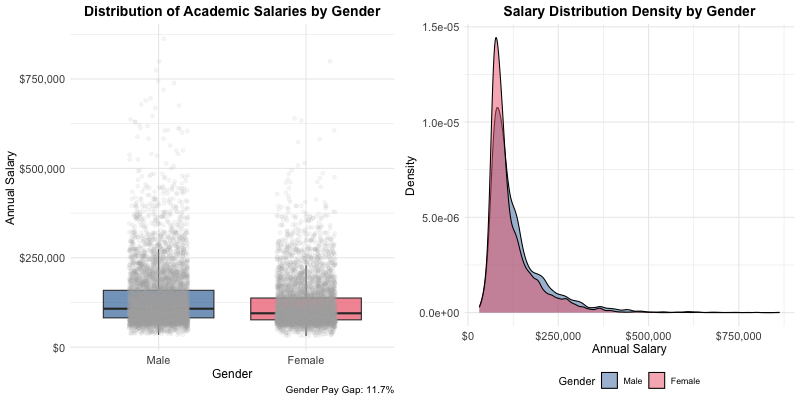}
   \caption{Distribution of Academic Salaries by Gender. Left panel shows the boxplot with individual observations (jittered). Right panel shows the density distribution of salaries by gender.}
   \label{fig:salary_dist2}
\end{figure}

In the full dataset, the results of a standard Ordinary Least Squares (OLS) regression model (Table~\ref{appendix:ols_regression}) indicate that the gender coefficient remains statistically significant even though considering professors without matched Google Scholar profiles, with $\beta = -0.0400$ ($SE = 0.0027$, $p < 2e^{-16}$). This corresponds to an estimated gender wage gap of 8.80\%. 
\label{appendix:ols_regression}
\begin{table}[H]
    \centering
    \renewcommand{\arraystretch}{1.2}
    \begin{tabular}{lcccc}
        \hline
        \textbf{Variable} & \textbf{Estimate} & \textbf{Std. Error} & \textbf{t-value} & \textbf{p-value} \\
        \hline
        (Intercept) & 4.7709 & 0.0060 & 792.959 & $<$2e-16*** \\
        Gender (Female) & -0.0400 & 0.0027 & -14.875 & $<$2e-16*** \\
        Google Scholar ID (Yes) & -0.0014 & 0.0027 & -0.524 & 0.6005 \\
        Titles: Associate Professor & 0.0654 & 0.0034 & 19.358 & $<$2e-16*** \\
        Titles: Professor & 0.1812 & 0.0040 & 44.783 & $<$2e-16*** \\
        University Code (DRU(H)) & 0.0534 & 0.0040 & 13.409 & $<$2e-16*** \\
        University Code (DU/VA) & 0.1653 & 0.0041 & 39.944 & $<$2e-16*** \\
        Department Code (B) & 0.2603 & 0.0058 & 44.737 & $<$2e-16*** \\
        Department Code (MHS) & 0.2836 & 0.0041 & 69.441 & $<$2e-16*** \\
        Department Code (NS) & 0.0715 & 0.0045 & 15.923 & $<$2e-16*** \\
        Department Code (SS) & 0.0720 & 0.0043 & 16.663 & $<$2e-16*** \\
        Department Code (TE) & 0.0892 & 0.0046 & 19.461 & $<$2e-16*** \\
        Working Years & -0.0021 & 0.0002 & -12.545 & $<$2e-16*** \\
        log$_{10}$(i10-index) & 0.0125 & 0.0033 & 3.794 & 0.000149*** \\
        \hline
    \end{tabular}
    \caption{OLS Regression Results: Log-Transformed Salary as Dependent Variable}
\end{table}

\subsection{Parametric Causal Analysis}

For the full dataset, we included Google Scholar ID as an additional covariate. We reported results for two estimands: the average treatment effect on the treated (ATT) and the average treatment effect (ATE).

\subsubsection{ATT: Popensity Score Matching}

\begin{figure}[H]
  \centering
  \includegraphics[width=0.75\textwidth]{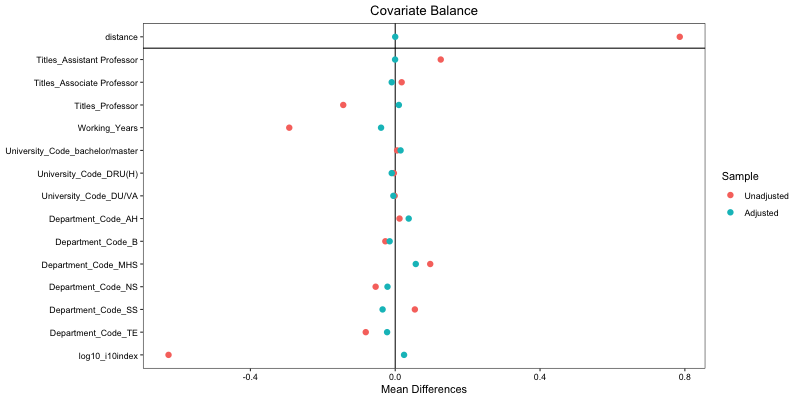}
  \caption{Love Plot for Full Dataset}
  \label{fig:love_plot2}
\end{figure}

Matching achieves excellent covariate balance (Table~\ref{tab:balance_measures}). The Love Plot (shown in Figure~\ref{fig:love_plot2}), which presents standardized mean differences (SMD) before and after matching, also illustrates a well-balanced covariate. The gender coefficient remains significant post-matching, with $\beta = -0.0269$ ($SE = 0.0040$, $p = 1.69e^{-11}$) (see Table~\ref{tab:interaction_model_final} for detailed information), corresponding to an estimated wage gap of 6.01\%. 

\label{appendix:matching_dia}

\begin{table}[H]
    \centering
    \renewcommand{\arraystretch}{1.2}
    \begin{tabular}{lccc}
        \hline
        \textbf{Covariate} & \textbf{Type} & \textbf{Diff. Unadjusted} & \textbf{Diff. Adjusted} \\
        \hline
        Distance & Distance & 0.786 & -0.000 \\
        Titles: Assistant Professor & Binary & 0.126 & -0.000 \\
        Titles: Associate Professor & Binary & 0.018 & -0.010 \\
        Titles: Professor & Binary & -0.144 & 0.010 \\
        Working Years & Continuous & -0.293 & -0.039 \\
        University Code: Bachelor/Master & Binary & 0.005 & 0.015 \\
        University Code: DRU(H) & Binary & -0.004 & -0.010 \\
        University Code: DU/VA & Binary & -0.002 & -0.005 \\
        Department Code: AH & Binary & 0.012 & 0.037 \\
        Department Code: B & Binary & -0.027 & -0.015 \\
        Department Code: MHS & Binary & 0.097 & 0.057 \\
        Department Code: NS & Binary & -0.054 & -0.021 \\
        Department Code: SS & Binary & 0.054 & -0.035 \\
        Department Code: TE & Binary & -0.081 & -0.022 \\
        $\log_{10}$(i10index) & Continuous & -0.626 & 0.024 \\
        \hline
    \end{tabular}
    \caption{Covariate Balance Before and After Matching}
    \label{tab:balance_measures}
\end{table}

\label{app:full_regression}
\begin{table}[H]
    \centering
    \renewcommand{\arraystretch}{1.2}
    \begin{tabular}{lcccc}
        \hline
        \textbf{Variable} & \textbf{Estimate} & \textbf{Std. Error} & \textbf{t-value} & \textbf{p-value} \\
        \hline
        (Intercept) & 5.0186 & 0.0076 & 664.39 & $<$2e-16*** \\
        Gender (Female) & -0.0269 & 0.0040 & -6.74 & 1.7e-11*** \\
        Google Scholar ID (Yes) & 0.0022 & 0.0037 & 0.61 & 0.5413 \\
        Assistant Prof $\times$ Working Years & -0.0071 & 0.0004 & -17.98 & $<$2e-16*** \\
        Associate Prof $\times$ Working Years & -0.0017 & 0.0003 & -6.45 & 1.2e-10*** \\
        Professor $\times$ Working Years & 0.0026 & 0.0002 & 13.10 & $<$2e-16*** \\
        bachelor/master $\times$ AH $\times$ log$_{10}$(i10-index) & -0.0995 & 0.0093 & -10.65 & $<$2e-16*** \\
        DRU(H) $\times$ AH $\times$ log$_{10}$(i10-index) & -0.0921 & 0.0067 & -13.69 & $<$2e-16*** \\
        DU/VA $\times$ AH $\times$ log$_{10}$(i10-index) & -0.0552 & 0.0076 & -7.27 & 4.1e-13*** \\
        bachelor/master $\times$ B $\times$ log$_{10}$(i10-index) & 0.0324 & 0.0149 & 2.18 & 0.0296* \\
        DRU(H) $\times$ B $\times$ log$_{10}$(i10-index) & 0.1081 & 0.0088 & 12.25 & $<$2e-16*** \\
        DU/VA $\times$ B $\times$ log$_{10}$(i10-index) & 0.1760 & 0.0102 & 17.22 & $<$2e-16*** \\
        bachelor/master $\times$ MHS $\times$ log$_{10}$(i10-index) & -0.0308 & 0.0119 & -2.58 & 0.0099** \\
        DRU(H) $\times$ MHS $\times$ log$_{10}$(i10-index) & 0.1030 & 0.0062 & 16.60 & $<$2e-16*** \\
        DU/VA $\times$ MHS $\times$ log$_{10}$(i10-index) & 0.1758 & 0.0046 & 38.01 & $<$2e-16*** \\
        bachelor/master $\times$ NS $\times$ log$_{10}$(i10-index) & -0.0739 & 0.0109 & -6.77 & 1.4e-11*** \\
        DRU(H) $\times$ NS $\times$ log$_{10}$(i10-index) & -0.0491 & 0.0072 & -6.87 & 7.0e-12*** \\
        DU/VA $\times$ NS $\times$ log$_{10}$(i10-index) & 0.0445 & 0.0053 & 8.44 & $<$2e-16*** \\
        bachelor/master $\times$ SS $\times$ log$_{10}$(i10-index) & -0.0776 & 0.0095 & -8.15 & 4.4e-16*** \\
        DRU(H) $\times$ SS $\times$ log$_{10}$(i10-index) & -0.0632 & 0.0068 & -9.33 & $<$2e-16*** \\
        DU/VA $\times$ SS $\times$ log$_{10}$(i10-index) & 0.0464 & 0.0060 & 7.66 & 2.0e-14*** \\
        bachelor/master $\times$ TE $\times$ log$_{10}$(i10-index) & -0.0757 & 0.0146 & -5.22 & 1.8e-07*** \\
        DRU(H) $\times$ TE $\times$ log$_{10}$(i10-index) & -0.0126 & 0.0077 & -1.65 & 0.0986. \\
        DU/VA $\times$ TE $\times$ log$_{10}$(i10-index) & 0.0453 & 0.0054 & 8.43 & $<$2e-16*** \\
        \hline
    \end{tabular}
    \caption{Three-Way Interaction Regression: Gender, Rank $\times$ Experience, and Productivity $\times$ Context Effects}
    \label{tab:interaction_model_final}
\end{table}

\subsubsection{ATE: IPTW and Regression Adjustment}
We estimate the average treatment effect (ATE) using inverse probability of treatment weighting (IPTW) and regression adjustment based on the propensity score.

The IPTW estimator yields a gender coefficient of $\beta = -0.0367$ (95\% CI: $[-0.0420, -0.0314]$), corresponding to an approximate 8.10\% salary gap. Regression adjustment produces a similar estimate of $\beta = -0.0369$ (95\% CI: $[-0.0448, -0.0291]$), corresponding to an 8.15\% gap. The estimates are consistent across methods.

\subsection{Non-Parametric Causal Analysis}

\begin{figure}[H]
    \centering
    \includegraphics[width=0.8\textwidth]{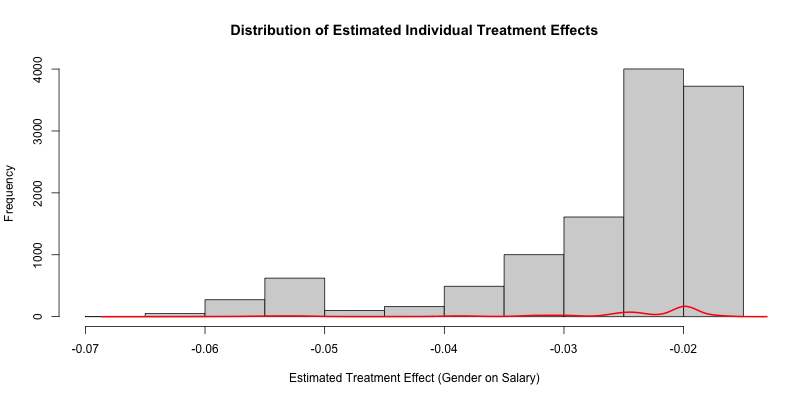}
    \caption{Distribution of Individual Treatment Effects}
    \label{fig:ite_dist2}
\end{figure}

\begin{figure}[H] 
    \centering
    \includegraphics[width=0.75\textwidth]{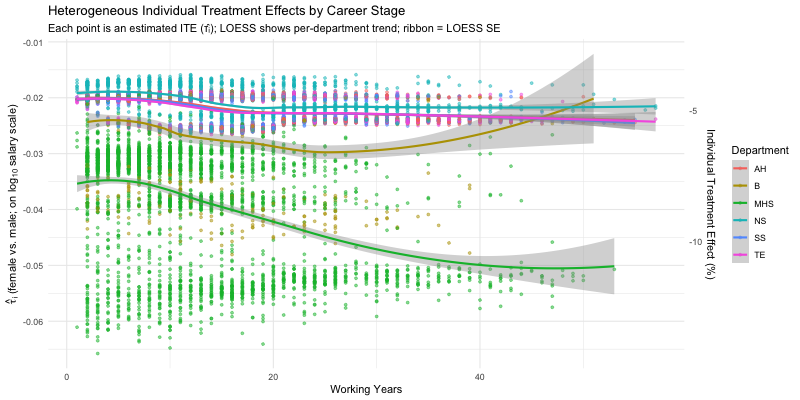}
    \caption{Heterogeneous Treatment Effects by Working Years.}
    \label{fig:working_years2}
\end{figure}

In the full dataset, the non-parametric Causal Forest model yields an Average Treatment Effect (ATE) of $-0.0262$ ($SE = 0.0036$), consistent with previous estimates. The Individual Treatment Effects (ITEs), shown in Figure~\ref{fig:ite_dist2}, range from $-0.066$ to $-0.016$, confirming considerable heterogeneity in gender-based salary disparities. Subgroup analyses (Figure~\ref{fig:working_years2}) indicate a trend toward improved gender pay equity with increasing years of experience.

\begin{figure}[H]
    \centering
    \includegraphics[width=0.75\textwidth]{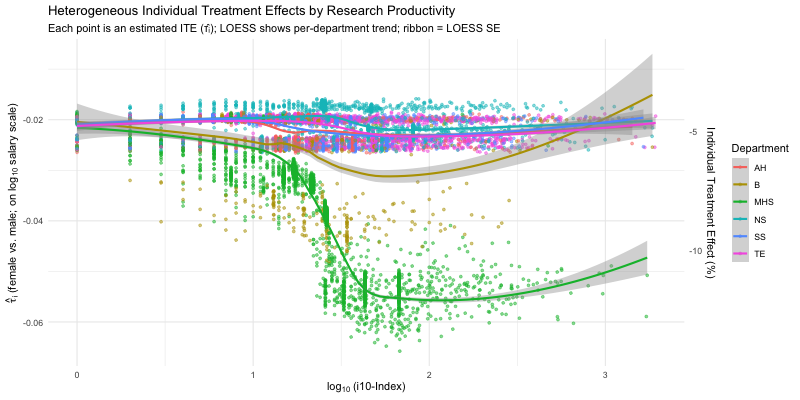}
    \caption{Treatment Effects by Research Productivity ($\log_{10}$ i10-index).}
    \label{fig:research_impact}
\end{figure}

\subsection{Summary of Comprehensive Dataset Results}

Overall, the results from the full dataset consistently indicate a substantial gender wage gap across all estimation methods. The unadjusted analysis suggests a gap of 12.80\%, which decreases after adjusting for observed covariates.

Across causal inference approaches, the estimated gender gap ranges from approximately 6\% to 8\% (Table~\ref{tab:summary_table2}). Propensity score matching (PSM) and causal forest yield slightly smaller estimates (around 6\%), while IPTW, regression adjustment, and OLS regression produce larger estimates (around 8\%). Despite these differences, all methods consistently indicate a statistically significant disadvantage for female faculty.

\begin{table}[H]
    \centering
    \caption{Estimated Gender Pay Gaps ($\log_{10}$ salary scale) by Method for Full Dataset. Estimates are reported with 95\% confidence intervals.}
    \label{tab:summary_table2}
    \begin{tabular}{llcc}
    \toprule
    \textbf{Method} & \textbf{Estimate $\beta$ (95\% CI)} & \textbf{Estimated Gap (\%)} \\
    \midrule
    Unadjusted Analysis & \textemdash{} & 12.80\% \\
    Baseline OLS Regression & -0.0400 (-0.0453, -0.0348) & 8.80\% (7.70\%, 9.91\%) \\
    OLS with Interaction Terms & -0.0372 (-0.0427, -0.0316) & 8.21\% (7.02\%, 9.36\%) \\
    PSM & -0.0269 (-0.0347, -0.0191) & 6.01\% (4.30\%, 7.68\%) \\
    IPTW & -0.0367 (-0.0420, -0.0314) & 8.10\% (6.97\%, 9.22\%)\\
    Regression Adjustment & -0.0369 (-0.0448, -0.0291) & 8.15\% (6.48\%, 9.80\%)\\
    Causal Forest & -0.0262 (-0.0333, -0.0190) & 5.85\% (4.28\%, 7.38\%)\\
    \bottomrule
    \end{tabular}
\end{table}

\end{document}